\documentclass[pdftex,twocolumn,epjc3]{svjour3}          
\RequirePackage[T1]{fontenc}
\smartqed  
\RequirePackage{amsmath,amssymb,color,dsfont,enumerate,flushend,graphicx,mathptmx,multirow,subfigure}
\RequirePackage[numbers,sort&compress]{natbib}
\RequirePackage[colorlinks,citecolor=blue,urlcolor=blue,linkcolor=blue]{hyperref}
\hypersetup{bookmarks=true,unicode=true,pdftoolbar=true,pdfmenubar=true,
  pdffitwindow=false,pdfstartview={FitH},
  pdftitle={lnvWithoutVR - EPJC (2017) 77: 375 - [arXiv:1703.04669]},pdfauthor={R.~Ruiz},
  pdfsubject={Subject},pdfcreator={Creator},pdfproducer={Producer},
  pdfkeywords={Majorana Neutrinos}{Left Right Symmetric Model}{Hadron Colliders},
  pdfnewwindow=true,colorlinks=true,
  linkcolor=blue,citecolor=magenta,filecolor=magenta,urlcolor=cyan}
\journalname{Eur. Phys. J. C}

\newcommand{\GeV}{{\rm ~GeV}}
\newcommand{\TeV}{{\rm ~TeV}}

\newcommand{\invfb}{{\rm ~fb^{-1}}}
\newcommand{\invab}{{\rm ~ab^{-1}}}
\newcommand{\vareps}{\varepsilon}

\newcommand{\WR}{W_R}
\newcommand{\ZR}{Z_R}
\newcommand{\MW}{M_{W}}
\newcommand{\MWR}{M_{W_R}}

\newcommand{\mN}{m_N}
\newcommand{\rN}{r_N}
\newcommand{\kR}{\kappa_R}

\newcommand{\BR}[1]{{{\rm BR}\left(#1\right)}}
\newcommand{\Gam}[1]{\Gamma\left(#1\right)}
\def\bsp#1\esp{\begin{split}#1\end{split}}

\newcommand{\confirm}{\textcolor{black}}

\begin{document}
\title{Lepton Number Violation at Colliders\\ from Kinematically Inaccessible Gauge Bosons}
\author{Richard Ruiz\thanksref{e1,addy1}}
\institute{
Institute for Particle Physics Phenomenology {\rm(IPPP)}, Department of Physics, Durham University, Durham, DH1 3LE, UK\label{addy1}
}
\thankstext{e1}{Electronic address: richard.ruiz@durham.ac.uk}
\date{Received: X / Accepted: X}
\maketitle

\begin{abstract}
We reevaluate the necessity of $W_R$ gauge bosons being kinematically accessible to test the Left-Right Symmetric Model (LRSM) at hadron colliders.
In the limit that $W_R$ are too heavy, resonant production of sub-TeV Majorana neutrinos $N$ can still proceed
at the Large Hadron Collider (LHC) via the process 
$pp\rightarrow W_R^{\pm *}\rightarrow N \ell^\pm \to \ell^\pm \ell^\pm +nj $ if mediated by a far off-shell $W_R$.
Traditional searches strategies are insensitive to this regime as they rely on momenta of final states scaling with TeV-scale $M_{W_R}$.
For such situations, the process
is actually kinematically and topologically identical to the direct production (DP) process
$pp\rightarrow W_{\rm SM}^{\pm *} \rightarrow N \ell^\pm \rightarrow \ell^\pm \ell^\pm +nj$.
In this context, we reinterpret {$\sqrt{s}=8$ TeV} LHC constraints on DP rates for the minimal LRSM.
For ${m_N = 200-500}$ GeV and right-left coupling ratio $\kappa_R = g_R/g_L$, we find $(M_{W_R} / \kappa_R) > {1.1-1.8}$ TeV at 95\% CLs.
Expected sensitivities to DP at 14 (100) TeV are also recast:
with $\mathcal{L}=1~(10)$~ab$^{-1}$,
one can probe $(M_{W_R} / \kappa_R) < {7.9-8.9~(14-40)}$ TeV for $m_N = {100 - 700~(1200)}$ GeV,
well beyond the anticipated sensitivity of resonant $W_R$ searches.
Findings in terms of gauge invariant dimension-six operators with heavy $N$ are also reported.
\end{abstract}

\section{Introduction}\label{sec:intro}
The Left-Right Symmetric model (LRSM)~\cite{Pati:1974yy,Mohapatra:1974hk,Mohapatra:1974gc,Senjanovic:1975rk,Senjanovic:1978ev} 
remains one of the best motivated high-energy
completions of the Standard Model of Particle Physics (SM).
It ties together the Majorana nature of neutrinos, 
their tiny masses in comparison to the electroweak (EW) scale $v_{\rm EW}$,
and the chiral structure of EW interactions, seemingly disparate phenomena,
to the simultaneous breakdown of $(B-L)$ conservation and left-right parity invariance at a scale $v_R\gg v_{\rm EW}$.
Predicting a plethora of observations,
the model is readily testable at current and near-future experiments;
see ~\cite{Chen:2011de,Mohapatra:2014cja,Senjanovic:2016bya,Mohapatra:2016twe,Arkani-Hamed:2015vfh,Golling:2016gvc} and references therein.

\begin{figure*}[!t]
\begin{center}
\subfigure[]{\includegraphics[width=.42\textwidth]{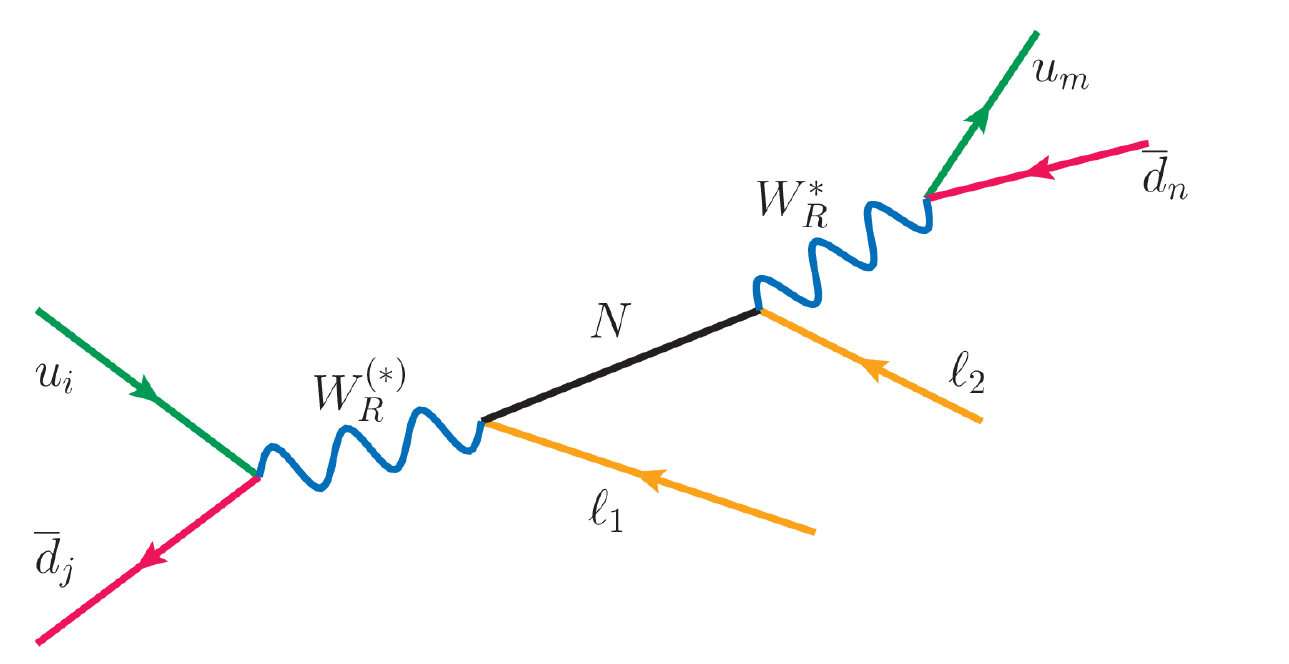}	\label{fig:feynman_LRSM_qqWR_Nl_NDecay}		}
\subfigure[]{\includegraphics[width=.42\textwidth]{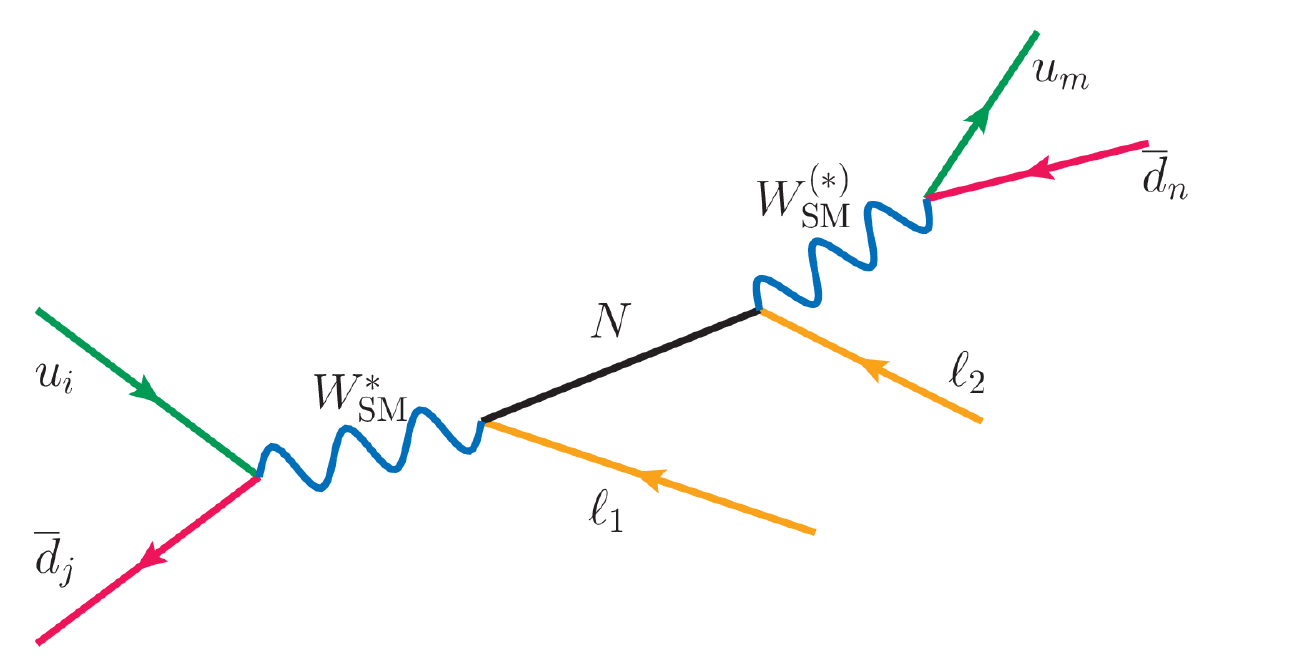}	\label{fig:feynman_Direct_qqWL_Nl_NDecay}	}
\end{center}
\caption{
Born diagrams for heavy Majorana $N$ production and decay via (a) $\WR$ (b) $W_{\rm SM}$ currents. Drawn using JaxoDraw~\cite{Binosi:2003yf}.}
\label{fig:feynmanBorn}
\end{figure*}

At the Large Hadron Collider (LHC), searches~\cite{Khachatryan:2014dka,Aad:2015xaa} for 
$\WR$ gauge bosons and heavy Majorana neutrinos $N$, if kinematically accessible,
focus on the well-studied, lepton number-violating 
$(\Delta L = \pm2)$ 
Drell-Yan process~\cite{Keung:1983uu},
\begin{equation}
  p ~p ~\rightarrow ~\WR^\pm ~\rightarrow N ~\ell^\pm_1 ~\rightarrow ~\ell^\pm_1 ~\ell^\pm_2 ~+nj.
  \label{eq:sslljjLRSM}
\end{equation}
As seen in Fig.~\ref{fig:feynman_LRSM_qqWR_Nl_NDecay}, Eq.~(\ref{eq:sslljjLRSM}) proceeds for $\mN<\MWR$
first through the on-shell production of $\WR$, then by its decay to $N$.
Recent investigations~\cite{Ferrari:2000sp,Maiezza:2015lza,Gluza:2016qqv,Mitra:2016kov,Mattelaer:2016ynf},
however, have shown that one can obtain a considerable increase in sensitivity to the LRSM at colliders 
by relaxing the requisite charged lepton and jet multiplicities stipulated by Ref.~\cite{Keung:1983uu} for Eq.~(\ref{eq:sslljjLRSM}) 
and similarly for the related single-top channel~\cite{Simmons:1996ws}. 
This is particularly true for $\MWR\gg\mN,~v_{\rm EW}$,
which occurs naturally when $v_R \gtrsim\mathcal{O}(10)$ TeV with neutrino triplet Yukawas $y^{\Delta_R} \lesssim \mathcal{O}(10^{-2})$.
Incidentally, such scenarios are also favored by searches for flavor-changing neutral Higgs (FCNH) 
transitions~\cite{Chakrabortty:2012pp,Bertolini:2014sua,Maiezza:2014ala,Maiezza:2016bzp} and neutron EDMs~\cite{Zhang:2007fn,Zhang:2007da}.
Along these lines, we reevaluate the necessity of $\WR$ being kinematically accessible to test LR symmetry at hadron colliders.

In the limit that $\MWR$ is of the order or above the total collider energy $\sqrt{s}$ but $\mN\ll\sqrt{s}$,
Eq.~(\ref{eq:sslljjLRSM}) can still proceed if mediated instead by a far \textit{off-shell} $\WR$.
This is akin to the SM Fermi contact interaction.
For $\mN\lesssim\mathcal{O}(1)\TeV$, 
8 TeV searches~\cite{Khachatryan:2014dka,Aad:2015xaa} for Eq.~(\ref{eq:sslljjLRSM})
are insensitive to this configuration due to the search premise itself:
resonant $\WR$ production implies that momenta of final-state particles scale with $\MWR$,
justifying the use of TeV-scale selection cuts in~\cite{Khachatryan:2014dka,Aad:2015xaa}.
The choice of cuts are motivated by limits from dijet searches that indicate $\MWR\gtrsim2.5\TeV$~\cite{Khachatryan:2015dcf,ATLAS:2015nsi}.
Non-resonant $\WR$ mediation, however, implies that the partonic scale is naturally $\sqrt{\hat{s}}\sim m_N\lesssim\mathcal{O}(1)\TeV$,
and therefore is unlikely to lead to final states satisfying the kinematical criteria.
For $m_N\gtrsim\mathcal{O}(1)$ TeV, present methods are sufficient~\cite{ATLAS:2012ak}.

Interestingly, while the underlying dynamics differ, for the $(\MWR,\mN)$ range in consideration, 
the mass scale and topology of Eq.~(\ref{eq:sslljjLRSM}) are identical to the heavy Majorana neutrino direct production (DP) process 
\begin{equation}
 p ~p ~\to ~W_{\rm SM}^{\pm *} ~\to ~\ell^\pm_1 ~N ~\to ~\ell^\pm_1 ~\ell^\pm_2 ~+nj.
 \label{eq:sslljjDirect}
\end{equation}
As shown in Fig.~\ref{fig:feynman_Direct_qqWL_Nl_NDecay}, this process,
which may also be labeled as prompt production, transpires through off-shell SM $W$ bosons
and occurs at the scale $\mN$ for $\mN > M_{W_{\rm SM}}$~\cite{Dicus:1991fk,Pilaftsis:1991ug,Datta:1993nm,Han:2006ip,Atre:2009rg}.
Subsequently, hadron collider searches for Eq.~(\ref{eq:sslljjDirect})
can be interpreted as searches for Eq.~(\ref{eq:sslljjLRSM}) in the $\MWR\gtrsim\sqrt{s}$ limit.
Moreover, despite its off-shell nature, the $\WR$ chiral couplings to quark and leptons remain
encoded in azimuthal and polar distributions of the $\ell^\pm\ell^\pm nj$ system~\cite{Han:2012vk}.
Thus, in principle, the dynamics of Eq.~(\ref{eq:sslljjLRSM}) can still be determined,
even in mixed $\WR^{(*)}-W_{\rm SM}^{(*)}$ scenarios as considered in~\cite{Han:2012vk,Chen:2013fna,Dev:2015kca}.
It follows that this holds too for $ee/pp\rightarrow \ZR^{(*)}\rightarrow NN$

In the LRSM, heavy $N$ production can in principle also proceed through Eq.~(\ref{eq:sslljjDirect}) and its neutral current equivalent via neutrino mixing.
However, such mixing between left-handed flavor states $\ell$ and heavy mass eigenstate $N$,
which scales as $V_{\ell N} \sim \sqrt{m_\nu / m_N}$, is necessarily small for the choice of $m_N$ in discussion and observed $m_\nu$.
Subsequently, we neglect the contribution of Eq.~(\ref{eq:sslljjDirect}) in the LRSM throughout this study.
For further discussions, see, e.g., Refs.~\cite{Nemevsek:2012iq,Chen:2013fna,Senjanovic:2016vxw}.

In this context, we reinterpret $\sqrt{s}=8$ TeV LHC limits 
on heavy Majorana neutrino DP cross sections~\cite{Khachatryan:2015gha,Khachatryan:2016olu} for the LRSM.
For $\confirm{m_N = 200-500}$ GeV and right-left coupling ratio $\kappa_R = g_R/g_L$,
we find $(M_{W_R} / \kappa_R) < \confirm{1.1-1.8}$ TeV are excluded at 95\% CLs.
While weak, the limits are competitive with searches for resonant $\MWR$-$N$ production~\cite{Aad:2015xaa,ATLAS:2012ak};
however, for such low mass scales, the validity of this approach requires $\kR\gg1$.
Projected sensitivities~\cite{Alva:2014gxa} to DP at the high-luminosity LHC 
and a hypothetical 100 TeV Very Large Hadron Collider (VLHC) are recast into projections for the LRSM.
At 14~(100) TeV and with $\mathcal{L}=1~(10)~\text{ab}^{-1}$,
one can probe $(M_{W_R} / \kappa_R) < \confirm{7.9-8.9~(14-40)}$ TeV for $m_N = \confirm{100 - 700~(1200)}$ GeV.
We also translate sensitivity to $(M_{W_R} / \kappa_R)$ for coefficients of gauge invariant dimension -six operators
in an Effective Field Theory with right-handed neutrinos (NEFT)~\cite{delAguila:2008ir}.

This study continues in the following order:
In Sec.~\ref{sec:model}, the components of LRSM and NEFT relevant for this work are reviewed.
We describe our methodology for reinterpreting (V)LHC limits in Sec.~\ref{sec:method},
and report results in Sec.~\ref{sec:results}.
We summarize and conclude in Sec.~\ref{sec:summary}.

\section{Theoretical Framework}\label{sec:model}
We now briefly summarize the main relations of the minimal LRSM and NEFT relevant to this analysis.

\subsection{Minimal Left-Right Symmetric Model}

In the notation of~\cite{Han:2012vk}, $\WR$ quark chiral currents are
\begin{eqnarray}
 \mathcal{L}_{\WR-q-q'} = \frac{-\kappa_R^q g_L}{\sqrt{2}}\sum_{i,j=u,\dots}\overline{u}_i V_{ij}^{R}~W_{R \mu}^+ \gamma^\mu P_R~ d_j + \text{H.c.}
 \nonumber
\end{eqnarray}
Here, up-(down-)type quarks with flavor $i (j)$ are represented by $u_i (d_j)$;
$P_{R(L)} = \frac{1}{2}(1\pm\gamma^5)$ is the right-hand [RH] (left-hand [LH]) chiral projection operator;
$V_{ij}^{\rm R}$ denotes the RH analog of Cabbibo-Kobayashi-Masakawa (CKM) matrix $V_{ij}^{\rm L}$; and
$\kappa_{R}^{q}\in\mathds{R}$ is an overall normalization for the $\WR$ interaction strength
with respect to the SM weak coupling $g_L=\sqrt{4\pi\alpha_{\rm EM}}/\sin\theta_W$.
Despite nature maximally violating parity at low energies, $V_{ij}^{\rm R}$ retains its resemblance to $V_{ij}^{\rm L}$,
with $\vert V_{ij}^{\rm R} \vert = \vert V_{ij}^{\rm L}\vert$ for generalized charge conjugation 
and $\vert V_{ij}^{\rm R} \vert \approx \vert V_{ij}^{\rm L}\vert+ \mathcal{O}(m_b/m_t)$ for 
generalized parity~\cite{Zhang:2007fn,Zhang:2007da,Maiezza:2010ic,Senjanovic:2014pva,Senjanovic:2015yea}.
Throughout this study, we assume five massless quarks and, for simplicity, take $\vert V^L_{ij}\vert,~\vert V^R_{ij}\vert$ to be diagonal with unit entries.

For leptonic coupling to $\WR$, we consider first the decomposition of neutrino chiral states $i,j$ into mass states $m,m'$:
Assuming  $i~(m) =1,\dots,3$, LH (light) states and $j~(m')=1,\dots,n$, RH (heavy) states,
we can relate chiral neutrino states and mass eigenstates by the rotation
\begin{eqnarray}
\begin{pmatrix} \nu_{Li} \\  N_{Rj}^c \end{pmatrix}
=
\begin{pmatrix} 
U_{3\times3} && V_{3\times n} \\
X_{n\times3} && Y_{n\times n}
\end{pmatrix}
\begin{pmatrix} \nu_{m} \\ N_{m'}^c \end{pmatrix}.
\label{eq:nuMixing}
\end{eqnarray}
Without the loss of generality, we take the rotation of the charged leptons into the mass basis as the identity.
The $U_{3 \times3}$ component of Eq.~(\ref{eq:nuMixing}) is then recognized as the observed light neutrino mixing matrix.
In analogy to $U_{\ell m}$, the entry $Y_{\ell m'} (X_{\ell m})$ quantifies the mixing between the heavy (light) mass state $N_{m'}~(\nu_{m})$  
and the RH chiral state with corresponding flavor $\ell$.
Hence, the mixing entries scale as
$\vert Y_{\ell m'}\vert^2 \sim \mathcal{O}(1)$ and
$\vert X_{\ell m}\vert^2 \sim 1 - \vert Y_{\ell m'}\vert^2 \sim \mathcal{O}(m_{\nu_m}/m_{N_{m'}})$~\cite{Keung:1983uu}.
Explicitly, the RH flavor state $N_{\ell}$ in the mass basis is then~\cite{Atre:2009rg,Han:2012vk}, 
\begin{equation}
 N_{\ell} = \sum_{m=1}^{3} X_{\ell m}\nu_{m}^c + \sum_{m'=1}^{n}Y_{\ell m'} N_{m'}.
 \label{eq:nuRDecomp}
\end{equation}
With this, the $\WR$ chiral currents for leptons are~\cite{Atre:2009rg,Han:2012vk}
\begin{eqnarray}
\mathcal{L}_{\WR-\ell-\nu/N} &=& \frac{-\kappa_R^\ell g_L}{\sqrt{2}}
 \sum_{\ell=e}^{\tau}
 \overline{N_{\ell}}~W_{R \mu}^+ \gamma^\mu P_R~ \ell^-+\text{H.c.}
 \nonumber\\
 &=&
 \frac{-\kappa_R^\ell g_L}{\sqrt{2}}
 \sum_{\ell=e}^{\tau}
 \Bigg[
 \sum_{m=1}^3	\overline{\nu^{c}_m} X_{\ell m}^\dagger +
 \sum_{m'=1}^3	\overline{N_{m'}} Y_{\ell m'}^\dagger
  \Bigg] 
 \nonumber\\
 & &~\times  ~W_{R \mu}^+ \gamma^\mu P_R~ \ell^-
 +\text{H.c.}
 \nonumber
\end{eqnarray}
As for quarks, $\kappa_R^\ell\in\mathds{R}$ normalizes the $\WR$ coupling to leptons.
Throughout this analysis, we adopt the conventional benchmark scenario
and consider only the lightest heavy neutrino mass state $N_{m'=1}$, which we denote as $N$.

\subsection{Effective Field Theory with Heavy Neutrinos}
Heavy Neutrino Effective Field Theory (NEFT)~\cite{delAguila:2008ir,Aparici:2009fh,Bhattacharya:2015vja} 
is a powerful extension of the SM EFT~\cite{Buchmuller:1985jz,Grzadkowski:2010es}
that allows for a consistent and agnostic parameterization of new, high-scale, 
weakly coupled physics when $N$ mass scales comparable to $v_{\rm EW}$.
As TeV-scale $L$ violation implies~\cite{Ma:1998dn,Kersten:2007vk} the existence of a particle spectrum 
beyond the canonical Type I seesaw~\cite{Minkowski:1977sc,GellMann:1980vs,Yanagida:1979as,Mohapatra:1979ia},
it is natural to consider DP sensitivities in terms of NEFT operators.

After extending the SM by three $N_R$, the most general renormalizable theory that can be constructed
from SM symmetries is the Type I Seesaw Lagrangian,
\begin{equation}
 \mathcal{L}_{\rm Type~I} = \mathcal{L}_{\rm SM} + \mathcal{L}_{N~\text{Kin.+Mass}} + \mathcal{L}_{N~\text{Yukawa}}.
\end{equation}
Respectively, the three terms are  the SM Lagrangian, the kinetic and Majorana mass terms for $N_R$, 
and the Yukawa couplings responsible for Dirac neutrino masses.
From this, the NEFT Lagrangian can be built by further extending $\mathcal{L}_{\rm Type~I}$
before EW symmetry breaking (EWSB) by all SU$(3)$ $\otimes$ SU$(2)_L$ $\otimes$ U$(1)_Y$-invariant, irrelevant (mass dimension $d>4$) operators
containing Type I Seesaw fields:
\begin{eqnarray}
 \mathcal{L}_{\rm NEFT} &=& \mathcal{L}_{\rm Type~I} + \sum_{d=5}\sum_{i} \frac{\alpha_i}{\Lambda^{(d-4)}}\mathcal{O}_{i}^{(d)}.
\end{eqnarray}
Here, $\alpha_i<\mathcal{O}(4\pi)$ are dimensionless coupling coefficients,
$\Lambda\gg\sqrt{\hat{s}}$ is the mass scale of the underlying theory,
and $\mathcal{O}_{i}^{(d)}$ are gauge invariant permutations of Type I field operators.
The list of $\mathcal{O}_{i}^{(d)}$ are known explicitly for $d=5$~\cite{Aparici:2009fh}, 
6~\cite{delAguila:2008ir}, and 7~\cite{Bhattacharya:2015vja},
and can be built for larger $d$  following~\cite{Henning:2015alf,Kobach:2016ami}.

At $d=6$, the four-fermion $\mathcal{O}_i^{(6)}$ giving rise to the same parametric
dependence on $\mN$ in the partonic cross section $\hat{\sigma}$
as both DP and the LRSM for $\MWR\gg\sqrt{\hat{s}}$ are
\begin{eqnarray}
 \mathcal{O}_V^{(6)} 	&=& \left(\overline{d}\gamma^\mu P_R u\right)\left(\overline{e}\gamma_\mu P_R N_R\right)
 \quad\text{and}\quad \nonumber\\
 \mathcal{O}_{S3}^{(6)} &=& \left(\overline{Q}\gamma^\mu P_R N_R\right)\varepsilon\left(\overline{L}\gamma_\mu P_R d\right).
 \label{eq:neftD6Ops}
\end{eqnarray}
In Eq.~(\ref{eq:neftD6Ops}), $\vareps$ is the totally antisymmetric tensor.
After EWSB and decomposing $N_R$ according to Eq.~(\ref{eq:nuRDecomp}),
but neglecting $\mathcal{O}(X_{\ell m})$ terms, the operators become
\begin{eqnarray}
  \mathcal{O}_V^{(6)} 	&=& \sum_{m'=1}\left(\overline{d}\gamma^\mu P_R u\right)\left(\overline{\ell}\gamma_\mu P_R~Y_{\ell m'}~N_{m'}\right)
 \quad\text{and}\quad\nonumber\\
 \mathcal{O}_{S3}^{(6)} &=& \sum_{m'=1}\left(\overline{Q}\gamma^\mu P_R ~Y_{\ell m'} N_{m'} \right)\left(\overline{\ell}\gamma_\mu P_R d\right).
 \label{eq:neftD6OpsMix}
\end{eqnarray}
As in the LRSM case, we consider only the $N_{m'=1}$ state with mixing as given in Eqs.~(\ref{eq:mumueeMix})-(\ref{eq:emuMix}).

\section{Mimicking Direction Production with Left-Right Symmetry}\label{sec:method}
In this section we describe our procedure for extracting bounds on LRSM and NEFT quantities 
from observed and expected (V)LHC limits on heavy Majorana neutrino DP rates.
Our computational setup is summarized in Sec.~(\ref{sec:setup}).
We start by constructing the observable $\vareps(\MWR)$, which we will ultimately constrain.

The Born-level, partonic heavy $N$ production cross section via (on- or off-shell) $\WR$ currents,
\begin{equation}
 q_1\overline{q_2} \to \WR^{\pm (*)} \to N ~\ell^\pm_1,
\end{equation}
with arbitrary lepton mixing is given generically by~\cite{Han:2012vk}
\begin{eqnarray}
 \frac{d\hat{\sigma}^{\rm LRSM}}{d\Omega_\ell} 
 = \frac{3\hat{\sigma}^{\rm LRSM}_{\rm Tot.}}{2^3\pi(2+\rN)}\left[(1-\cos\theta_\ell)^2  + \rN\sin^2\theta_\ell\right]
 \label{eq:lrsmDXSec}
\end{eqnarray}
where $\rN \equiv \mN^2/\hat{s}$ and the total cross section is
\begin{eqnarray}
  \hat{\sigma}^{\rm LRSM}_{\rm Tot.} &=& 
  \cfrac{\kR^{q2}\kR^{\ell2} g_L^4}{2^7~3N_c~\pi}
  \cfrac{\vert Y_{\ell N}\vert^2~\hat{s}(1-\rN)^2(2+\rN)}{\left[(\hat{s}-\MWR^2)^2 + (\MWR \Gamma_{W_R})^2\right]} ~
  \\
  &\approx&
  \cfrac{\kR^{q2}\kR^{\ell2} g_L^4 }{2^7~3N_c~\pi}\vert Y_{\ell N}\vert^2
  \cfrac{\hat{s}}{\MWR^4}(1-\rN)^2(2+\rN).
   \label{eq:lrsmTotXSec}
\end{eqnarray}
In the last line we take the $\MWR\gg\sqrt{\hat{s}}$ limit.
For DP, the analogous partonic cross section is
\begin{eqnarray}
 \frac{d\hat{\sigma}^{\rm DP}}{d\Omega_\ell} 
 = \frac{3\hat{\sigma}^{\rm DP}_{\rm Tot.}}{2^3\pi(2+\rN)}\left[(1-\cos\theta_\ell)^2  + \rN\sin^2\theta_\ell\right]
  \label{eq:dpDXSec}
\end{eqnarray}
where the total partonic rate for $\sqrt{\hat{s}}\gg M_{W_{\rm SM}}$ is similarly,
\begin{eqnarray}
  \hat{\sigma}^{\rm DP}_{\rm Tot.} &=& 
  \cfrac{g_L^4}{2^7~3N_c~\pi}
  \cfrac{\vert V_{\ell N}\vert^2~\hat{s}(1-\rN)^2(2+\rN)}{\left[(\hat{s}-\MWR^2)^2 + (\MW \Gamma_{W})^2\right]} ~
  \label{eq:dpTotXSecFull}
  \\
  &\approx&
  \cfrac{g_L^4 \vert V_{\ell N}\vert^2}{2^7~3N_c~\pi}
  \cfrac{1}{\hat{s}}(1-\rN)^2(2+\rN)
   \label{eq:dpTotXSec}
\end{eqnarray}

Comparing the differential and integrated expressions
one sees crucially that the angular and $\mN$ dependence in the two processes are the same.
This follows from the maximally parity violating $V\pm A$ structures of the $W_{\rm SM}/\WR$ couplings.
Na\"ively, one expects the orthogonal chiral couplings to invert the leptons' polarizations with respect to the mediator.
However, as the mediators' polarizations are also relatively flipped with respect to the initial-state quarks, 
the outgoing lepton polarization with respect to initial-state quarks, i.e., $\cos\theta_\ell$, is the same.
Hence, universality of $\WR$ chiral couplings to quarks and leptons in the LRSM can be tested without resonantly producing it.
The precise handedness of the couplings can be inferred from azimuthal and polar distributions 
of the $\ell^\pm\ell^\pm j j$ final state~\cite{Han:2012vk} as well as single-top channel~\cite{Gopalakrishna:2010xm}.
As DP searches do not (and should not) rely on forward-backward cuts, 
which are sensitive to parity asymmetries, their reinterpretation in terms of the LRSM for non-resonant $\WR$ is justified.

Branching rates of $N$ to a final state $A$ can be expressed in terms of the calculable $N\rightarrow A$ partial widths,
\begin{equation}
 \BR{N\rightarrow A} \equiv \cfrac{\Gam{N\rightarrow A}}{\sum_i ~\Gam{N\rightarrow A_i}}.
 \label{eq:brDef}
\end{equation}
For $\MWR\gg\mN$, the $\MWR$ dependence in Eq.~(\ref{eq:brDef}) cancels.
Hence, the Born-level, partonic same-sign lepton cross section in the LRSM,
\begin{equation}
 q_1\overline{q_2} \to \WR^{\pm*} \to N ~\ell^\pm_1 \to \ell^\pm_1 ~\ell^\pm_2 ~X,
 \label{eq:sslljjProcess}
\end{equation}
under the narrow width approximation for $N$ is
\begin{eqnarray}
\hat{\sigma}(q_1\overline{q_2}  &\to& N ~\ell^\pm_1 \to \ell^\pm_1 ~\ell^\pm_2 ~X)
\nonumber\\
&\approx&   \hat{\sigma}^{\rm LRSM}_{\rm Tot.} ~\times~ \BR{N \to \ell^\pm_2 ~X} 
 \\
 &\equiv& \vareps^{\ell_1\ell_2}(\MWR) ~\times~ \tilde{\hat{\sigma}}.
 \label{eq:epsRedXSecDef}
\end{eqnarray}
In the last line we collect LRSM parameters into the single, dimensionful [TeV$^{-4}$] coefficient 
\begin{eqnarray}
 \vareps^{\ell_1\ell_2}(\MWR) =   \cfrac{\kR^{q 2}\kR^{\ell 2}}{\MWR^4}\vert Y_{\ell_1 N}\vert^2 ~  \BR{N \to ~\ell^\pm_2 ~q'_1\overline{q'_2}}.
 \quad
 \label{eq:epsDef}
\end{eqnarray}
The ``reduced'' partonic cross section $\tilde{\hat{\sigma}}$ contains all kinematical and $m_N$ dependence that 
must be convolved with parton distribution functions (PDFs) to build the hadronic cross section. 
For the $e^\pm\mu^\pm$ mixed-flavor state, a summation over $ \vareps^{e\mu}$ and $\vareps^{\mu e}$ is implied.

Inclusive, hadronic level cross sections are obtained from the Collinear Factorization Theorem,
\begin{eqnarray}
& & \sigma(pp \to A+X) = f \otimes f \otimes \hat{\sigma} 
\\
& & = \frac{1}{\delta_{ij}+1} \sum_{i,j=u,g,\dots}\int^1_{\tau_0} d\xi_1 \int^1_{\tau_0/\xi_1} d\xi_2  
\nonumber\\
&  & \Big[f_{i/p}(\xi_1,\mu_f)f_{j/p}(\xi_2,\mu_f) + (1\leftrightarrow2)\Big]\hat{\sigma}(ij\rightarrow A).
\end{eqnarray}
It expresses the production rate of $A$ (and arbitrary beam remnant $X$) in $pp$ collisions
as the convolution $(\otimes)$ of the $ij\rightarrow A$ partonic process rate
and the process-independent PDFs $f_{k/p}(\xi,\mu_f)$,
which for parton species $k$ with longitudinal momentum  $p_z = \xi E_p$ resums collinear splittings up to the scale $\mu_f$.
The kinematic threshold $\tau_0$ is the scale below which the process is kinematically forbidden.
For heavy $N$ production, $\tau_0 = \mN^2 / s$.
In terms of $\vareps(\MWR)$, the hadronic equivalent of Eq.~(\ref{eq:epsRedXSecDef}) is
\begin{eqnarray}
  \sigma(p ~p ~\to N ~\ell^\pm_1 \to \ell^\pm_1 ~\ell^\pm_2 + X)
  =
  \vareps(\MWR) \times \tilde{\sigma}.
  \label{eq:loHadronicFactorization}
\end{eqnarray}
Here, $\tilde{\sigma}$ is the ``reduced'' hadronic cross section and is related to $\tilde{\hat{\sigma}}$ by 
the convolutions $\tilde{\sigma} =  f \otimes f \otimes \tilde{\hat{\sigma}}$.
As the next-to-leading order (NLO) in QCD corrections for arbitrary DY processes largely factorize 
from the hard scattering process~\cite{Harris:2001sx,Ruiz:2015zca}, Eq.~(\ref{eq:loHadronicFactorization}) holds at NLO:
\begin{eqnarray}
  \sigma^{\rm NLO}(p ~p \to N ~\ell^\pm_1 \to \ell^\pm_1 ~\ell^\pm_2 + X)  =  \vareps(\MWR) \times \tilde{\sigma^{\rm NLO}}.
  \nonumber\\
  \label{eq:nloHadronicFactorization}
\end{eqnarray}
Premising that reported LHC limits on the DP cross section can be applied to the LRSM for kinematically inaccessible $\WR$, 
Eq.~(\ref{eq:nloHadronicFactorization}) shows how to translate the upper bound on the rate into an upper bound on $\vareps(\MWR)$.

For the NEFT operators in Eq.~(\ref{eq:neftD6Ops}), the corresponding partonic scattering rates are given by~\cite{delAguila:2008ir}
\begin{eqnarray}
\hat{\sigma}_{S3}(u\overline{d}\to N\ell^\pm_1 \to \ell^\pm_1\ell^\pm_2 X) &=&
 \frac{\alpha_{S3}^2 \vert Y_N\ell_1\vert^2}{2^7~3N_c \pi} \frac{\hat{s}}{\Lambda^4}
 \nonumber\\
\times(1-\rN)^2(2+\rN)  &\times&  \BR{N \to \ell_2 X}, \qquad 
 \label{eq:eftXsecS3} \\ 
\hat{\sigma}_V(u\overline{d}\to N\ell^\pm_1 \to \ell^\pm_1\ell^\pm_2 X)    &=& 
 \frac{4\alpha_V^2}{\alpha_{S3}^2}\hat{\sigma}_{S3}.
 \label{eq:eftXsecV} 
\end{eqnarray}
Comparing to Eqs.~(\ref{eq:lrsmDXSec})-(\ref{eq:dpDXSec}), one finds the mapping
\begin{eqnarray}
 \mathcal{O}_{S3}^{(6)}	&:&	\vareps^{\ell_1\ell_2}(\MWR) = \frac{\alpha_{S3}^2}{\Lambda^4}\vert Y_N\ell_1\vert^2\BR{N\to \ell_2 X},~\qquad
 \label{eq:od6S3Map}
 \\
  \mathcal{O}_V^{(6)}	&:&	\vareps^{\ell_1\ell_2}(\MWR) = \frac{4\alpha_V^2}{\Lambda^4}\vert Y_N\ell_1\vert^2 \BR{N\to \ell_2 X}.~\qquad
 \label{eq:od6VMap}
\end{eqnarray}
and allows the further interpretation of $\vareps(\MWR)$.

\begin{figure*}[!t]
\begin{center}
\subfigure[]{\includegraphics[width=.44\textwidth]{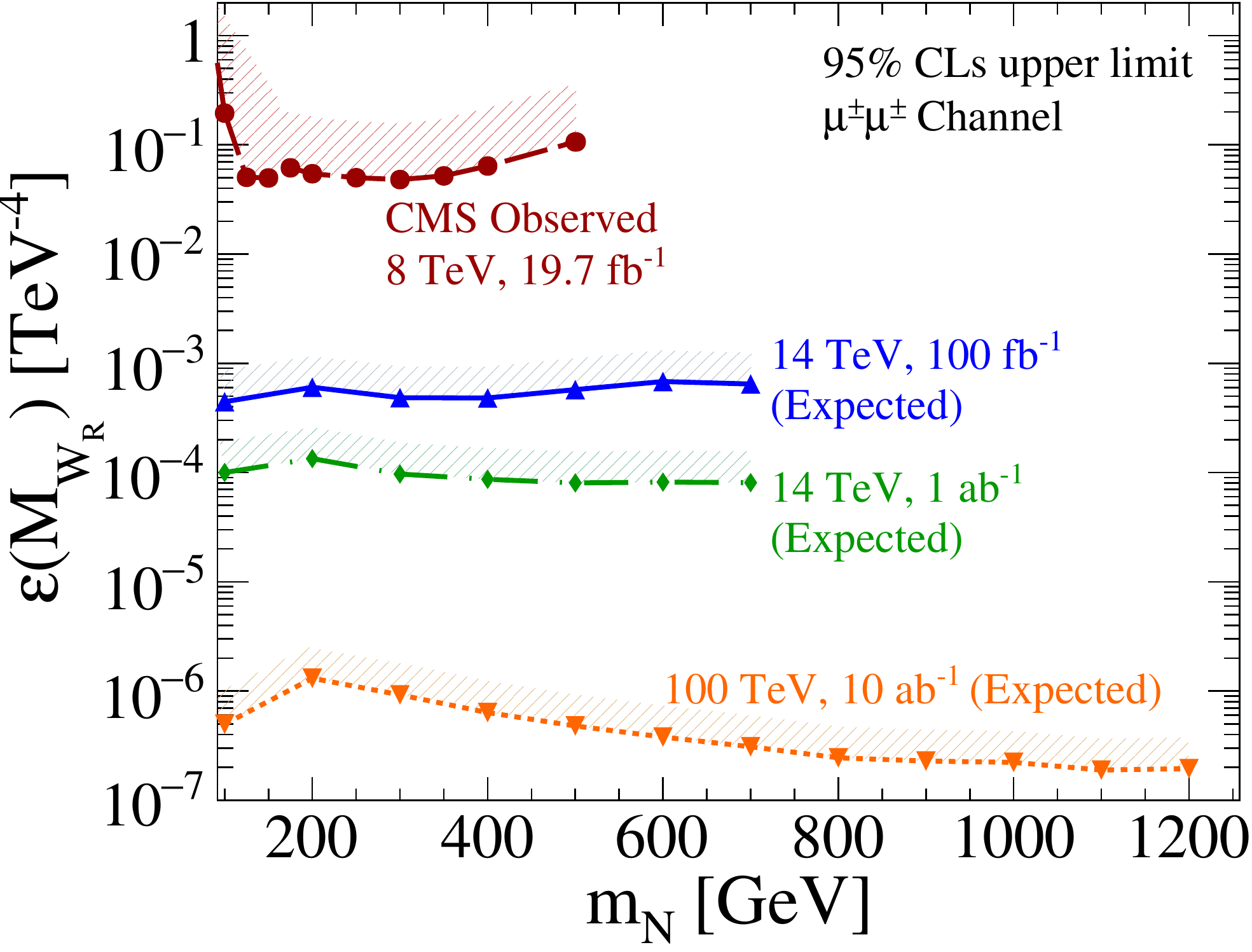}	\label{fig:lnvWithoutVR_epsMWR_dimuon}	}
\subfigure[]{\includegraphics[width=.44\textwidth]{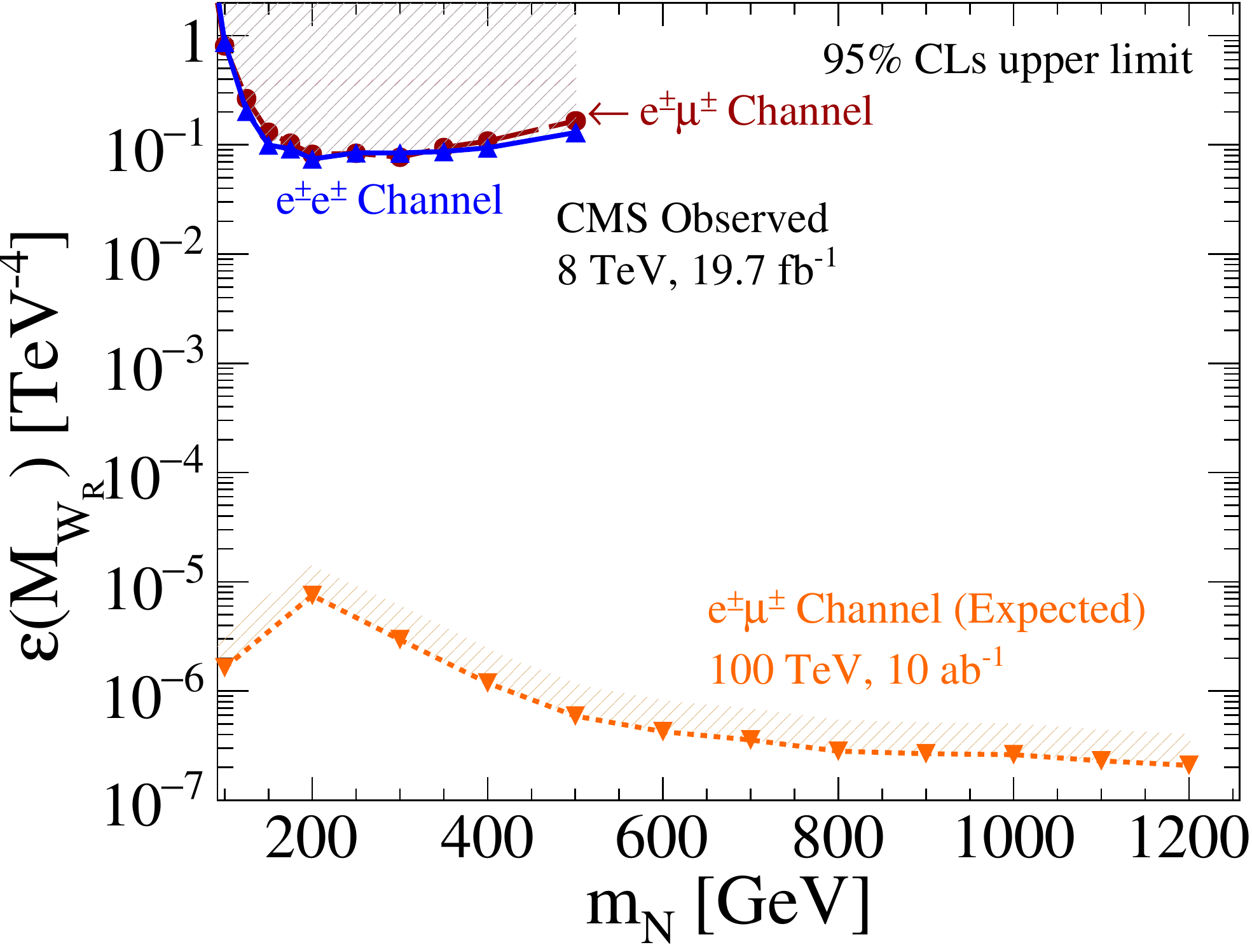}	\label{fig:lnvWithoutVR_epsMWR_emuXee}	}
\\
\subfigure[]{\includegraphics[width=.44\textwidth]{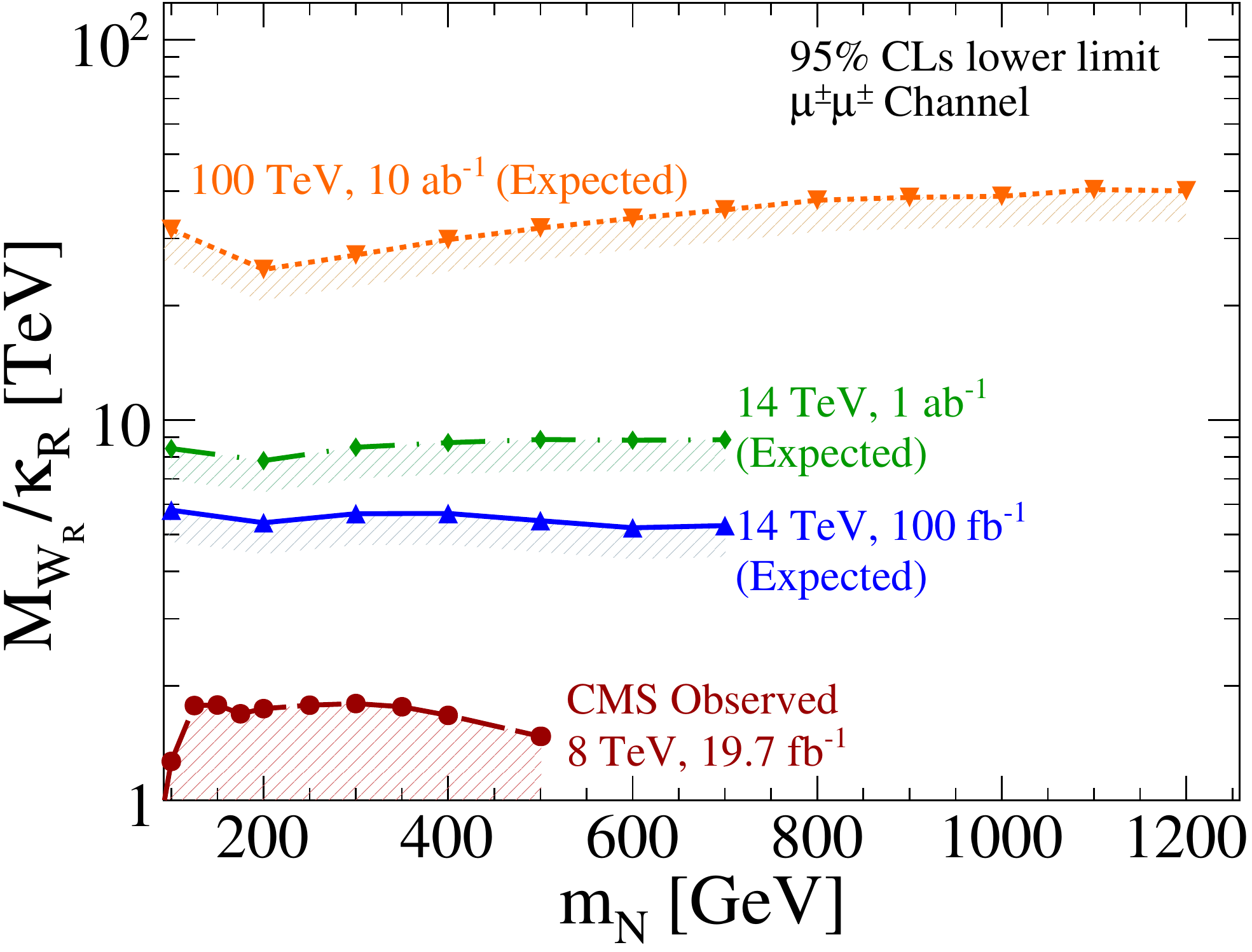}	\label{fig:lnvWithoutVR_mwrKapR_dimuon}	}
\subfigure[]{\includegraphics[width=.44\textwidth]{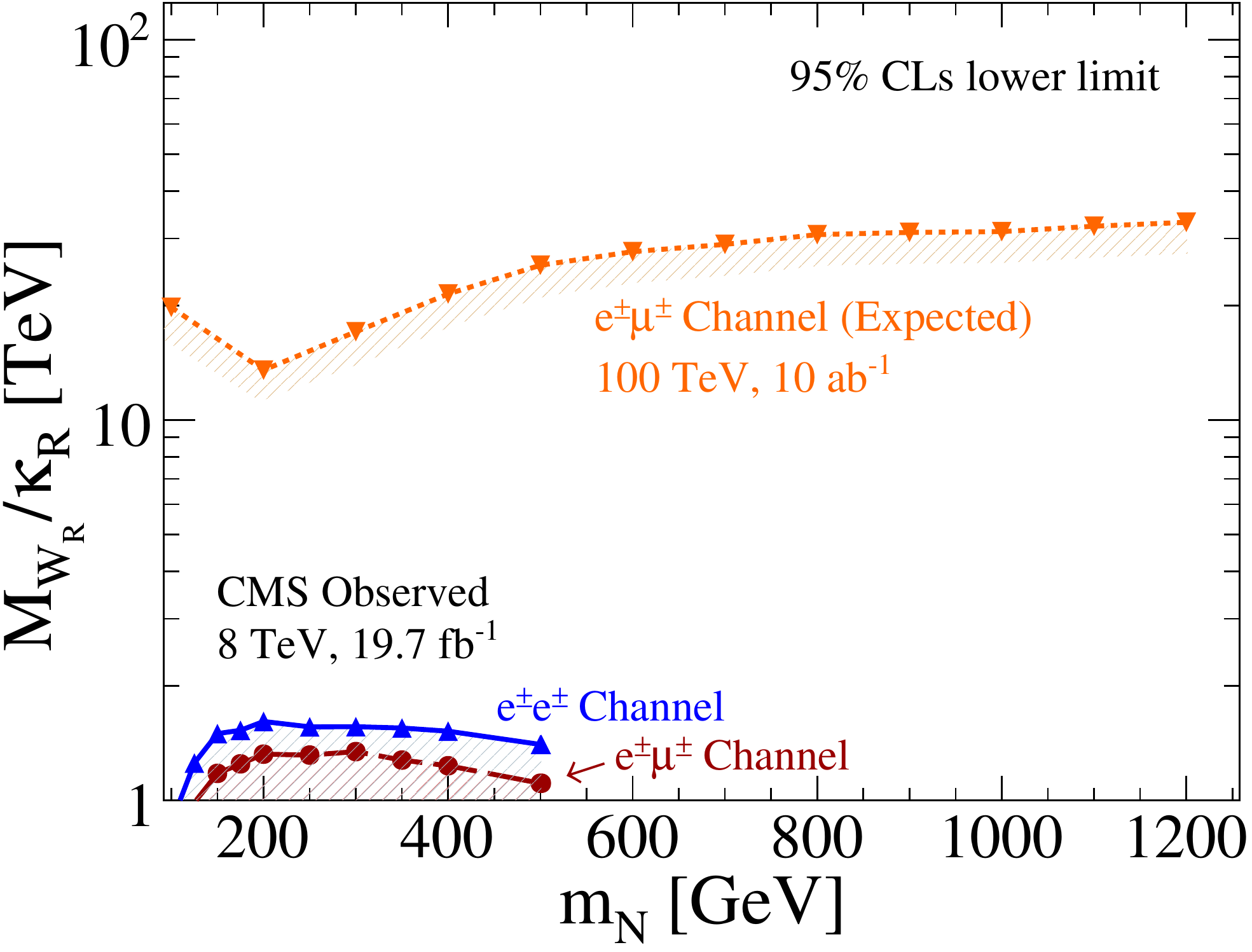}	\label{fig:lnvWithoutVR_mwrKapR_emuXee}	}
\end{center}
\caption{
(a) As a function of $\mN$, observed 8 TeV LHC upper bound on $\vareps^{\mu\mu}(\MWR)$ (dash-dot),
expected 14 TeV sensitivity with $\mathcal{L}=100\invfb$ (solid-triangle) and $1\invab$ (dash-dot-diamond),
and expected 100 TeV VLHC sensitivity with $10\invab$ (\confirm{dot-star}).
(b) Same as (a) but with $e^\pm\mu^\pm$ (dash-dot) and $e^\pm e^\pm$ (solid-triangle) at 8 TeV
and $e\mu$ (\confirm{dot-star}) at 100 TeV.
(c,d) Same as (a,b), respectively, but for lower bounds on $(\MWR/\kR)$.
All limits are obtained at 95\% CL$_s$.
}
\label{fig:lrsmMimicry_Limits}
\end{figure*}

\subsection{Computational Setup}\label{sec:setup}
Practically speaking, the NLO-accurate reduced cross section is determined using 
the FeynRules-based~\cite{Christensen:2008py,Alloul:2013bka,Degrande:2014vpa}
NLO-accurate \texttt{Effective Left-Right Symmetric Model} file of~\cite{Mattelaer:2016ynf} and MadGraph5\_amc@NLO~\cite{Alwall:2014hca}.
The processes,
\begin{equation}
 p p \to \WR^{\pm *} \to N \mu^\pm ~+X
\end{equation}
is calculated at NLO accuracy assuming test inputs:
\begin{eqnarray}
\{M_{\rm Test}\} &:& \MWR = \confirm{200}\TeV, ~\kR^{\ell,q}=1,\nonumber\\
		 & &  \vert Y_{\mu N} \vert = 1,  ~\BR{N\rightarrow\mu X}=1.\quad
 \label{eq:lrsmTestInputs}
\end{eqnarray}
For choice of EW inputs, PDFs, etc., we follow Ref.~\cite{Mattelaer:2016ynf}.
Denoting the $\vareps(\MWR)$ corresponding to the Eq.~(\ref{eq:lrsmTestInputs}) as $\vareps(M_{\rm Test})$, 
$\tilde{\sigma}^{\rm NLO}$ is obtained from 
the relationship
\begin{equation}
 \tilde{\sigma}^{\rm NLO} = \frac{\sigma^{\rm NLO}(p ~p ~\to  N ~\mu^\pm + X; \{M_{\rm Test}\})}{\vareps(M_{\rm Test})}.
 \label{eq:reducedNLO}
\end{equation}

\section{Results and Discussion}\label{sec:results}
\begin{figure*}[!t]
\begin{center}
\subfigure[]{\includegraphics[width=.45\textwidth,height=6.3cm]{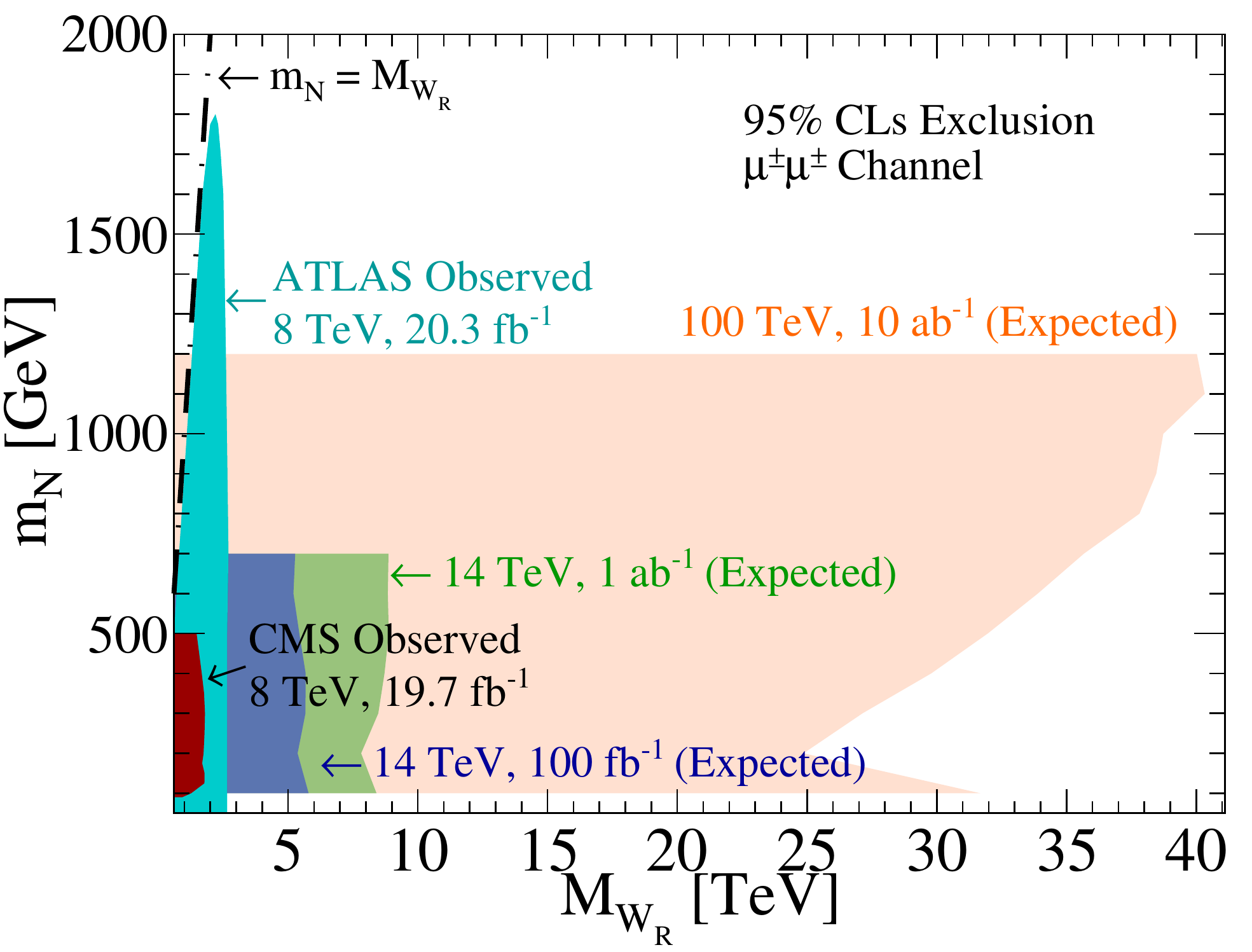}	\label{fig:lnvWithoutVR_MWRmNExcl}	}
\subfigure[]{\includegraphics[width=.45\textwidth]{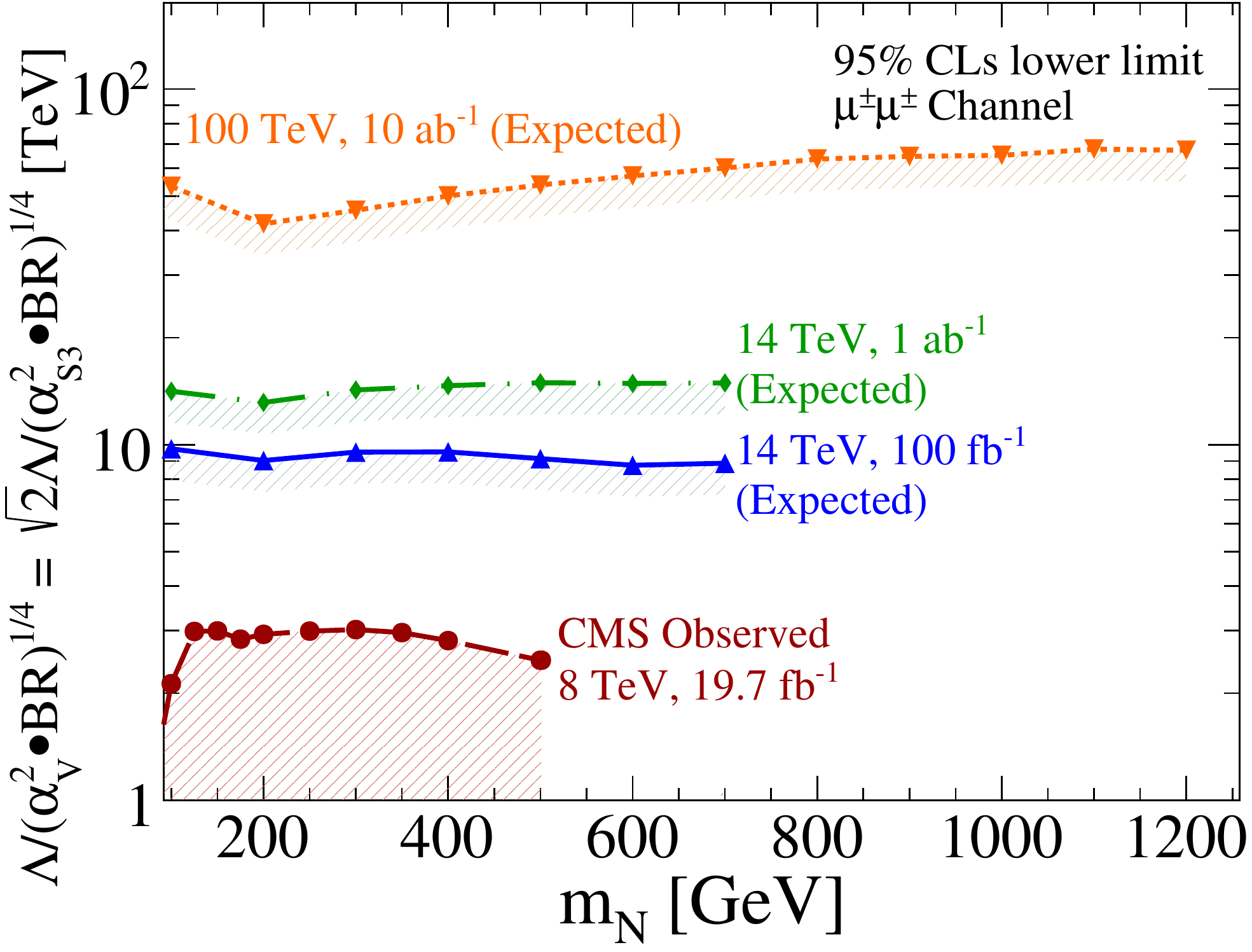}		\label{fig:lrsmMimicry_NEFT}		}
\end{center}
\caption{
(a) Observed and expected 95\% CL$_s$ sensitivities to the $(\MWR,\mN)$ parameter space $(\kR=1)$ for various collider configurations via
direct and indirect searches in the $\mu^\pm\mu^\pm$ final state.
(b) Observed and expected 95\% CL$_s$ sensitivities to the NEFT dimension-six operators $\mathcal{O}^{(6)}_V$ and $\mathcal{O}^{(6)}_{S3}$
in the $\mu^\pm\mu^\pm$ channel for the collider configurations in Fig.~\ref{fig:lnvWithoutVR_epsMWR_dimuon}.
}
\label{fig:lrsmMimicry_recast}
\end{figure*}

We now report the observed sensitivity to the LRSM  from DP searches in the  $\mu\mu/ee/e\mu$ channels
by the CMS experiment at $\sqrt{s}=8$ TeV with $\mathcal{L}=19.7\invfb$~\cite{Khachatryan:2015gha,Khachatryan:2016olu}.
We also report expected sensitivities based on 14 TeV projections with  $\mathcal{L}=100\invfb$ and $1\invab$~\cite{Alva:2014gxa},
as well as at 100 TeV with $\mathcal{L}=10\invab$~\cite{Alva:2014gxa}.
In all cases, 95\% confidence level (CL) limits are obtained/reproduced 
via the CL$_s$ method~\cite{Read:2002hq,Junk:1999kv,ATLAS:2011tau}, 
using the information available in~\cite{Khachatryan:2015gha,Khachatryan:2016olu,Alva:2014gxa},
and assuming Poisson distributions for signal and background processes.
After obtaining the expected (observed) DP cross section limits $\sigma^{{\rm 95\% CL}_s}_{\rm Exp.~(Obs.)}$, 
LRSM constraints are determined from the ``reduced'' cross section $\tilde{\sigma}$, as defined in Eq.~(\ref{eq:reducedNLO}), 
with the relation
\begin{equation}
 \vareps^{\ell_1\ell_2}_{\rm Exp.~(Obs.)}(\MWR) = \cfrac{\sigma^{{\rm 95\% CL}_s}_{\rm Exp.~(Obs.)}}{\tilde{\sigma}^{\rm NLO}}.
\end{equation}

In Fig.~\ref{fig:lrsmMimicry_Limits} we plot as a function of $m_N$ the 8 TeV CMS upper bounds on $\vareps(\MWR)$
for the (a) $\mu\mu$ (dash-dot) as well as (b) $e\mu$ (dash-dot) and $ee$ (\confirm{upside-down triangle}) channels.
One finds comparable limits for all modes, with
\begin{eqnarray}
  \mu^\pm\mu^\pm		&:&	\vareps^{\ell\ell}(\MWR) \lesssim \confirm{0.05}\TeV^{-4}, \\
  e^\pm\mu^\pm,~e^\pm e^\pm	&:&  	\vareps^{\ell\ell}(\MWR) \lesssim \confirm{0.1}\TeV^{-4}.
\end{eqnarray}
For $\mN\lesssim150\GeV$, $W_{\rm SM}$ production greatly diminishes sensitivity.
A weaker limit for $e$-based channels is due to the larger fake and charge misidentification rates for electrons than for muons,
particularly from top quarks.
These features are seen consistently in projections.

In Fig.~\ref{fig:lnvWithoutVR_epsMWR_dimuon}, the expected sensitivity to $\vareps^{\mu\mu}(\MWR)$ 
at 14 TeV with $\mathcal{L}=100\invfb$ (solid-triangle) and $1\invab$ (dash-dot-diamond) are shown.
We find that for $\mN = 100-700\GeV$, one can potentially exclude:
\begin{eqnarray}
 \mathcal{L}^{14\TeV}_{100\invfb}	&:&	\vareps^{\mu\mu}(\MWR) \lesssim \confirm{5\times10^{-4}\TeV^{-4}}, \\
 \mathcal{L}^{14\TeV}_{1\invab}		&:&	\vareps^{\mu\mu}(\MWR) \lesssim \confirm{9\times10^{-5}\TeV^{-4}}. 
\end{eqnarray}

At a future 100 TeV VLHC, the large increase in parton density coupled with proposed integrated luminosity goals
of 10-20$\invab$~\cite{Hinchliffe:2015qma} implies a considerable jump in sensitivity to $\vareps(\MWR)$ for EW-scale $N$.
For $\mN=100-1200\GeV$, the $\mu\mu$ (\confirm{dot-star}) in \ref{fig:lnvWithoutVR_epsMWR_dimuon}) and 
$e\mu$ (\confirm{dot-star}) in \ref{fig:lnvWithoutVR_epsMWR_emuXee}) final state can probe with $10\invab$:
\begin{eqnarray}
 \vareps^{\mu\mu}(\MWR)	&\lesssim& \confirm{2\times10^{-7}-1\times10^{-6}\TeV^{-4}}, \\
 \vareps^{e\mu}(\MWR) 	&\lesssim& \confirm{2\times10^{-7}-7\times10^{-6}\TeV^{-4}}.
\end{eqnarray}

\begin{table*}
\centering
\small
\caption{
Observed~\cite{Khachatryan:2015gha,Khachatryan:2016olu} and 
expected~\cite{Alva:2014gxa} 95\% CL$_s$ sensitivities to $\vareps(\MWR)$ and $(\MWR/\kR)$ in the LRSM as well as 
$\Lambda/\sqrt[4]{\alpha_V^2 \text{BR}}$ in NEFT assuming various $pp$ collider energies $(\sqrt{s})$ and integrated luminosity caches $(\mathcal{L})$.
} 

\begin{tabular}{ c || c || c | c | c | c || c | c | c | c || c | c | c | c }
\hline\hline
\multicolumn{2}{c||}{}	 & Obs.	&  \multicolumn{3}{c||}{Exp.} & Obs.	&  \multicolumn{3}{c||}{Exp.} & Obs.	&  \multicolumn{3}{c}{Exp.} \\ \hline
$\sqrt{s}$ [TeV]	 &	& 8 & 14 & 14 & 100 & 8 & 14 & 14 & 100 & 8 & 14 & 14 & 100 \\					
$\mathcal{L}$ [$\invfb$] &	& 19.7 & 100 & $10^3$ & $10^4$	& 19.7 & 100 & $10^3$ & $10^4$	& 19.7 & 100 & $10^3$ & $10^4$	\\ \hline\hline
$\mN$ [GeV]		& $\ell^\pm_1\ell^\pm_2$	& \multicolumn{4}{c|}{$\vareps(\MWR)$~[TeV$^{-4}$]}			 					 
							& \multicolumn{4}{c|}{$\MWR/\kR$ [TeV]}					
							& \multicolumn{4}{c}{$\Lambda/\sqrt[4]{\alpha_V^2 \cdot \text{BR}}$ [TeV]} 				
\\ \hline\hline
\multirow{3}{*}{100}	& $\mu\mu$	& $1.95\times10^{-1}$	& $6.45\times10^{-4}$	& $1.00\times10^{-4}$	& $4.96\times10^{-7}$ 	& 1.3		& 5.8		& 8.4		& 32		& 2.1		& 9.7		& 14		& 53 	\\ 
			& $e\mu$	& $8.05\times10^{-1}$	& --		& --		& $1.64\times10^{-6}$	& 0.75		& --	& --	& 20		& 1.5		& --	& --	& 40 	\\
			& $ee$		& $8.70\times10^{-1}$	& --		& --		& --		& 0.87		& --	& --	& --	& 1.5		& --	& --	& -- \\ \hline
\multirow{3}{*}{200}	& $\mu\mu$	& $5.44\times10^{-2}$	& $6.03\times10^{-4}$	& $1.34\times10^{-4}$	& $1.31\times10^{-6}$	& 1.7		& 5.4		& 7.8		& 25		& 2.9		& 9.0		& 13		& 42 	\\ 
			& $e\mu$	& $8.19\times10^{-2}$	& --		& --		& $7.49\times10^{-6}$	& 1.3		& --	& --	& 14		& 2.6		& --	& --	& 27 	\\
			& $ee$		& $7.42\times10^{-2}$	& --		& --		& --		& 1.6		& --	& --	& --	& 2.7		& --	& --	& -- \\ \hline
\multirow{3}{*}{300}	& $\mu\mu$	& $4.81\times10^{-2}$	& $6.84\times10^{-4}$	& $9.69\times10^{-5}$	& $9.22\times10^{-7}$	& 1.8		& 5.7		& 8.5		& 27		& 3.0		& 9.5		& 14		& 46 	\\ 
			& $e\mu$	& $7.70\times10^{-2}$	& --		& --		& $2.95\times10^{-6}$	& 1.3		& --	& --	& 17		& 2.7		& --	& --	& 34 	\\
			& $ee$		& $8.42\times10^{-2}$	& --		& --		& --		& 1.6		& --	& --	& --	& 2.6		& --	& --	& -- \\ \hline
\multirow{3}{*}{500}	& $\mu\mu$	& $1.06\times10^{-1}$	& $5.74\times10^{-4}$	& $8.04\times10^{-5}$	& $4.79\times10^{-7}$	& 1.5		& 5.4		& 8.9		& 32.		& 2.5		& 9.1		& 15		& 54 	\\ 
			& $e\mu$	& $1.66\times10^{-1}$	& --		& --		& $5.90\times10^{-7}$	& 1.1		& --	& --	& 26		& 2.2		& --	& --	& 51 	\\
			& $ee$		& $1.29\times10^{-1}$	& --		& --		& --		& 1.4		& --	& --	& --	& 2.4		& --	& --	& -- \\ \hline
\multirow{2}{*}{1200}	& $\mu\mu$	& --		& --		& --		& $1.95\times10^{-7}$	& --	& --	& --	& 40		& --	& --	& --	& 67 	\\ 
			& $e\mu$	& --		& --		& --		& $2.09\times10^{-7}$	& --	& --	& --	& 33		& --	& --	& --	& 66 	\\ \hline\hline
\end{tabular}
\label{tb:lrsmSensitivity}
\end{table*}

Derived limits on $\vareps(\MWR)$ hold for rather generic LR scenarios.
Under the strong (but typical) assumptions of a minimal LRSM setting, we can 
rewrite constraints as lower bounds on ratio of $\MWR$ and $\kR^{q,\ell}$.
Specifically, assuming gauge coupling universality, one has
\begin{equation}
 \kR \equiv \kR^{q} = \kR^{\ell}.
\end{equation}
For single flavor final-states, we take the aligned lepton mixing limit Eq.~(\ref{eq:mumueeMix}),
whereas for the mixed flavor channel, we take the maximally mixed limit Eq.~(\ref{eq:emuMix}), i.e.,
\begin{eqnarray}
 & & \vert Y_{\ell N}\vert \approx 1 \quad\text{and}\quad \BR{N\rightarrow \ell^\pm X}  \approx 1, \text{or}\qquad
 \label{eq:mumueeMix}
 \\
 & & \vert Y_{e N}\vert \approx \vert Y_{\mu N}\vert \approx 1/\sqrt{2}
 \quad\text{and}\quad \nonumber\\
 & & \BR{N \to e^\pm X} \approx \BR{N\rightarrow \mu^\pm X} \approx1/2.\qquad
  \label{eq:emuMix}
\end{eqnarray}
While $N$ can decay with equal 
likelihood to $\ell_i^+$ and $\ell_i^-$,
the same-sign charge stipulation reduces the effective branching by $1/2$.
With this, we invert $\vareps(\MWR)$, giving
\begin{equation}
 \frac{\MWR}{\kR} = \frac{1}{\sqrt[4]{\eta\times\vareps^{\ell_1\ell_2}(\MWR)}}, \quad 
 \eta = \left\{\begin{matrix}
2, & \ell_1 = \ell_2 \\
4, & \ell_1 \neq \ell_2
\end{matrix}\right.,
\end{equation}
where $\eta$ accounts for charge and flavor multiplicities.

In Figs.~\ref{fig:lnvWithoutVR_mwrKapR_dimuon} and ~\ref{fig:lnvWithoutVR_mwrKapR_emuXee}, respectively,
we show the lower bounds on $(\MWR/\kR)$ for the same configurations as (a) and (b).
For all channels, the observed 8 TeV limits span:
\begin{eqnarray}
 \mN=100-200\GeV	&:&	\left(\frac{\MWR}{\kR}\right) \gtrsim \confirm{0.7-1.8}\TeV, \nonumber\\
 \mN=200-700\GeV	&:&	\left(\frac{\MWR}{\kR}\right) \gtrsim \confirm{1.1-1.8}\TeV. \nonumber
\end{eqnarray}
At $\sqrt{s}=14$ TeV with $\mathcal{L}=100\invfb$ and $1\invab$, the $\mu\mu$ final state can exclude for $\mN = 100-700\GeV$:
\begin{eqnarray}
 \mathcal{L}^{14\TeV}_{100\invfb}	&:&	\left(\frac{\MWR}{\kR}\right) \lesssim \confirm{5.2-5.8}\TeV, \\
 \mathcal{L}^{14\TeV}_{1\invab}		&:&	\left(\frac{\MWR}{\kR}\right) \lesssim \confirm{7.8-8.9}\TeV. \quad
\end{eqnarray}
Comparable sensitivity in the $ee$ and $e\mu$ channels is expected.
At 100 TeV with $10\invab$, the $\mu\mu$ and $e\mu$ channels for $\mN=100-1200\GeV$ are sensitive to
\begin{eqnarray}
 \mu^\pm\mu^\pm	&:&	\left(\frac{\MWR}{\kR}\right) \lesssim \confirm{25-40}\TeV, \\
 e^\pm\mu^\pm	&:&	\left(\frac{\MWR}{\kR}\right) \lesssim \confirm{14-33}\TeV.
\end{eqnarray}
We note that the sharp cutoffs at $m_N=500,~700,~$ and ~1200 GeV for the several scenarios in Fig.~\ref{fig:lnvWithoutVR_MWRmNExcl}
is due to the limited number of mass hypotheses considered in~\cite{Khachatryan:2015gha,Khachatryan:2016olu,Alva:2014gxa}.
A dedicated analysis would show sensitivity to larger $m_N$.

To compare with searches for resonant $\WR$-$N$ production, we plot in Fig.~\ref{fig:lnvWithoutVR_MWRmNExcl} 
the region of the $(\MWR,\mN)$ parameter space 
excluded by the ATLAS experiment at 8 TeV with $20.3\invfb$~\confirm{in the $\mu\mu$ channel}~\cite{Aad:2015xaa},
along with our corresponding sensitivities for $\kR=1$.
For $\mN \approx 100-500\GeV,$ we find that the reinterpretation of CMS's DP limits are actually
within $\confirm{1.5\times}$ of present $\MWR$ limits from resonant $\WR$-$N$ and  dijet (not shown) 
searches~\cite{Aad:2015xaa,Khachatryan:2014dka,Khachatryan:2015dcf,ATLAS:2015nsi}.
However, for such low mass scales, the validity of this approach requires $\kR\gg1$.
With 100$\invfb$ at 14 TeV, projected sensitivities are competitive with 
the $\mathcal{O}(5)$ TeV reach from resonant searches using the full HL-LHC dataset~\cite{Ferrari:2000sp,Han:2012vk,Mitra:2016kov}.
With 1$\invab$ at 14 TeV, and more so with  10$\invab$ at 100 TeV,
the DP channel can probe  super heavy $v_R$ scales favored by 
low-energy probes~\cite{Bertolini:2014sua,Maiezza:2014ala,Maiezza:2016bzp,Zhang:2007fn,Zhang:2007da}.
These findings suggest searches for heavy Majorana neutrinos via off-shell $\WR$ may be of  some 
usefulness at current and future collider experiments.

For completeness, upper limits on $\vareps^{\mu\mu}(\MWR)$ are recast in terms of the NEFT operators in Eq.~(\ref{eq:neftD6OpsMix}).
Using Eqs.~(\ref{eq:od6VMap})-(\ref{eq:od6S3Map}), the lower bounds on $(\Lambda/\sqrt{\alpha_{V,S3}})$ are
\begin{eqnarray}
 \cfrac{\Lambda}{\sqrt[4]{\alpha_V^2 \BR{N\rightarrow\mu X}}}    &>& \sqrt[4]{\cfrac{4 \vert Y_{\mu N}\vert^2}{\vareps^{\mu\mu}_{\rm Exp~(Obs)}(\MWR)}},
 \\
 \cfrac{\Lambda}{\sqrt[4]{\alpha_{S3}^2 \BR{N\rightarrow\mu X}}} &>& \sqrt[4]{\cfrac{\vert Y_{\mu N}\vert^2}{\vareps^{\mu\mu}_{\rm Exp~(Obs)}(\MWR)}}.
\end{eqnarray}
As a function of $\mN$, the observed and expected sensitivities to $\mathcal{O}_V$ 
for the several configurations in Fig.~\ref{fig:lnvWithoutVR_epsMWR_dimuon} and mixing choice in Eq.~(\ref{eq:mumueeMix})
are shown in Fig.~\ref{fig:lrsmMimicry_NEFT}. Over the respective ranges of $\mN$, they span approximately
\begin{eqnarray}
 \mathcal{L}^{8\TeV}_{19.7\invfb}	
 &:& \frac{\Lambda}{\sqrt[4]{\alpha_V^2 \BR{N\rightarrow\mu X}}} > \confirm{2.1-3.0}\TeV, \quad \quad
 \\
 \mathcal{L}^{14\TeV}_{100\invfb}
 &:& \frac{\Lambda}{\sqrt[4]{\alpha_V^2 \BR{N\rightarrow\mu X}}} > \confirm{8.7-9.7}\TeV,
 \\
 \mathcal{L}^{14\TeV}_{1\invab}
 &:& \frac{\Lambda}{\sqrt[4]{\alpha_V^2 \BR{N\rightarrow\mu X}}} > \confirm{13-15}\TeV,
 \\
 \mathcal{L}^{100\TeV}_{10\invab}
 &:& \frac{\Lambda}{\sqrt[4]{\alpha_V^2 \BR{N\rightarrow\mu X}}} > \confirm{42-68}\TeV. 
\end{eqnarray}

We summarize our reported findings in Tbl.~\ref{tb:lrsmSensitivity}.

\section{Summary and Conclusion}\label{sec:summary}
While the LRSM naturally addresses shortcomings of the SM,
it is not guaranteed its entire particle spectrum lies within the kinematic reach of the LHC or a future 100 TeV VLHC.
Indeed, low-energy probes favor the LR breaking scale to be above the LHC's 
threshold~\cite{Chakrabortty:2012pp,Bertolini:2014sua,Maiezza:2014ala,Maiezza:2016bzp,Zhang:2007fn,Zhang:2007da}.

In this context, we argue that when LRSM gauge bosons are too heavy to be produced resonantly, 
on-shell production of sub-TeV Majorana neutrinos via the process $pp\to \WR^* \to N\ell^\pm \to \ell^\pm\ell^\pm + nj$ 
is still possible when mediated by far \textit{off-shell} $\WR$.
In this regime, the process' mass scale and topology are identical to the direct production (DP) 
process $pp\to W_{\rm SM}^{*} \to N\ell^\pm \to \ell^\pm\ell^\pm + nj$. 
Subsequently, searches for DP of heavy Majorana neutrinos can be translated into searches for LR symmetry.

We have recast current~\cite{Khachatryan:2014dka,Aad:2015xaa}
and projected~\cite{Han:2012vk,Alva:2014gxa} sensitivities to the DP process  at $pp$ colliders 
into observed and expected sensitivities for the LRSM, in the heavy $\MWR$ limit.
We find the following:
\begin{enumerate}[i)]
 \item At the 8 TeV LHC, for $\mN=100-500\GeV$ and right-left coupling ratio $\kappa_R = g_R/g_L$,
 searches  have excluded at 95\% CL$_s$ ${(\MWR/\kR)< 0.7-1.8\TeV}$.
 For $\mN\gtrsim200$ GeV, this is within ${1.5\times}$ of searches for resonant $\WR$ and $\WR$-$N$ production.
 
 \item At 14 TeV with $100\invfb~(1\invab)$, one can exclude at 95\% CL$_s$
 $\confirm{(\MWR/\kR) < 5.2-5.8~(7.8-8.9)}$ TeV
  for $\mN=100-700\GeV$, well beyond the $\mathcal{O}(5)$ TeV anticipated reach of resonant $\WR$ searches.
 
 \item At 100 TeV with $10\invab$, one can probe $(\MWR/\kR) < 14-40\TeV$ at 95\% CL$_s$ for $\mN=100-1200\GeV$,
 thereby greatly complimenting low-energy probes of $\mathcal{O}(10)$ TeV $v_R$. 
 
 \item In terms of an Effective Field Theory featuring heavy neutrinos, we find limits on mass/coupling scales
 for gauge invariant, dimension six operators  comparable to the aforementioned limits in the LRSM.
\end{enumerate}

\begin{acknowledgements}
Peter Ballett, Lydia Brenner, Luca Di Luzio, Silvia Pascoli, Carlos Fibo Tamarit, and Cedric Weiland are thanked for discussions. 
This work was funded in part by the UK Science and Technology Facilities Council, and
the European Union's Horizon 2020 research and innovation programme under the Marie Sklodowska-Curie grant agreement 674896 (Elusives ITN).
\end{acknowledgements}





\begin{thebibliography}{99} 
  \bibitem{Pati:1974yy} 
  J.~C.~Pati and A.~Salam,
  Phys.\ Rev.\ D {\bf 10}, 275 (1974)
  Erratum: [Phys.\ Rev.\ D {\bf 11}, 703 (1975)].
  doi:10.1103/PhysRevD.10.275, 10.1103/PhysRevD.11.703.2
  
  \bibitem{Mohapatra:1974hk} 
  R.~N.~Mohapatra and J.~C.~Pati,
  Phys.\ Rev.\ D {\bf 11}, 566 (1975).
  doi:10.1103/PhysRevD.11.566
  
  \bibitem{Mohapatra:1974gc} 
  R.~N.~Mohapatra and J.~C.~Pati,
  Phys.\ Rev.\ D {\bf 11}, 2558 (1975).
  doi:10.1103/PhysRevD.11.2558
  
  \bibitem{Senjanovic:1975rk} 
  G.~Senjanovic and R.~N.~Mohapatra,
  Phys.\ Rev.\ D {\bf 12}, 1502 (1975).
  doi:10.1103/PhysRevD.12.1502

  \bibitem{Senjanovic:1978ev} 
  G.~Senjanovic,
  Nucl.\ Phys.\ B {\bf 153}, 334 (1979).
  doi:10.1016/0550-3213(79)90604-7
  
  \bibitem{Arkani-Hamed:2015vfh} 
  N.~Arkani-Hamed, T.~Han, M.~Mangano and L.~T.~Wang,
  Phys.\ Rept.\  {\bf 652}, 1 (2016)
  doi:10.1016/j.physrep.2016.07.004
  [arXiv:1511.06495 [hep-ph]].
  
  \bibitem{Golling:2016gvc} 
  T.~Golling {\it et al.},
  [arXiv:1606.00947 [hep-ph]].
  
  
  \bibitem{Chen:2011de} 
  M.~C.~Chen and J.~Huang,
  Mod.\ Phys.\ Lett.\ A {\bf 26}, 1147 (2011)
  doi:10.1142/S0217732311035985
  [arXiv:1105.3188 [hep-ph]].
  
  \bibitem{Mohapatra:2014cja} 
  R.~N.~Mohapatra,
  PoS Neutel {\bf 2013}, 050 (2013).

  \bibitem{Mohapatra:2016twe} 
  R.~N.~Mohapatra,
  Nucl.\ Phys.\ B {\bf 908}, 423 (2016).
  doi:10.1016/j.nuclphysb.2016.03.006
  
  \bibitem{Senjanovic:2016bya} 
  G.~Senjanovic,
  Mod.\ Phys.\ Lett.\ A {\bf 32}, no. 04, 1730004 (2017)
  doi:10.1142/S021773231730004X
  [arXiv:1610.04209 [hep-ph]].
  
  
  \bibitem{Khachatryan:2014dka} 
  V.~Khachatryan {\it et al.} [CMS Collaboration],
  Eur.\ Phys.\ J.\ C {\bf 74}, no. 11, 3149 (2014)
  doi:10.1140/epjc/s10052-014-3149-z
  [arXiv:1407.3683 [hep-ex]].
  
  \bibitem{Aad:2015xaa} 
  G.~Aad {\it et al.} [ATLAS Collaboration],
  JHEP {\bf 1507}, 162 (2015)
  doi:10.1007/JHEP07(2015)162
  [arXiv:1506.06020 [hep-ex]].
  
  
  \bibitem{Keung:1983uu} 
  W.~Y.~Keung and G.~Senjanovic,
  Phys.\ Rev.\ Lett.\  {\bf 50}, 1427 (1983).
  doi:10.1103/PhysRevLett.50.1427
  
  \bibitem{Binosi:2003yf} 
  D.~Binosi and L.~Theussl,
  \textit{JaxoDraw: A Graphical user interface for drawing Feynman diagrams,}
  Comput.\ Phys.\ Commun.\  {\bf 161}, 76 (2004)
  [hep-ph/0309015].

  
  \bibitem{Ferrari:2000sp} 
  A.~Ferrari, J.~Collot, M.~L.~Andrieux, B.~Belhorma, P.~de Saintignon, J.~Y.~Hostachy, P.~Martin and M.~Wielers,
  Phys.\ Rev.\ D {\bf 62}, 013001 (2000).
  doi:10.1103/PhysRevD.62.013001
  
  \bibitem{Maiezza:2015lza} 
  A.~Maiezza, M.~Nemevšek and F.~Nesti,
  Phys.\ Rev.\ Lett.\  {\bf 115}, 081802 (2015)
  doi:10.1103/PhysRevLett.115.081802
  [arXiv:1503.06834 [hep-ph]].
  
  \bibitem{Gluza:2016qqv} 
  J.~Gluza, T.~Jelinski and R.~Szafron,
  Phys.\ Rev.\ D {\bf 93}, no. 11, 113017 (2016)
  doi:10.1103/PhysRevD.93.113017
  [arXiv:1604.01388 [hep-ph]].
 
 
  \bibitem{Mitra:2016kov} 
  M.~Mitra, R.~Ruiz, D.~J.~Scott and M.~Spannowsky,
  Phys.\ Rev.\ D {\bf 94}, no. 9, 095016 (2016)
  doi:10.1103/PhysRevD.94.095016
  [arXiv:1607.03504 [hep-ph]].
  
  \bibitem{Mattelaer:2016ynf} 
  O.~Mattelaer, M.~Mitra and R.~Ruiz,
  arXiv:1610.08985 [hep-ph].
    
  \bibitem{Simmons:1996ws} 
  E.~H.~Simmons,
  ``New gauge interactions and single top quark production,''
  Phys.\ Rev.\ D {\bf 55}, 5494 (1997)
  doi:10.1103/PhysRevD.55.5494
  [hep-ph/9612402]. 
  
  \bibitem{Chakrabortty:2012pp} 
  J.~Chakrabortty, J.~Gluza, R.~Sevillano and R.~Szafron,
  JHEP {\bf 1207}, 038 (2012)
  doi:10.1007/JHEP07(2012)038
  [arXiv:1204.0736 [hep-ph]].
  
  \bibitem{Bertolini:2014sua} 
  S.~Bertolini, A.~Maiezza and F.~Nesti,
  Phys.\ Rev.\ D {\bf 89}, no. 9, 095028 (2014)
  doi:10.1103/PhysRevD.89.095028
  [arXiv:1403.7112 [hep-ph]].
  
  \bibitem{Maiezza:2014ala} 
  A.~Maiezza and M.~Nemevšek,
  Phys.\ Rev.\ D {\bf 90}, no. 9, 095002 (2014)
  doi:10.1103/PhysRevD.90.095002
  [arXiv:1407.3678 [hep-ph]].
  
  \bibitem{Maiezza:2016bzp} 
  A.~Maiezza, M.~Nemevšek and F.~Nesti,
  Phys.\ Rev.\ D {\bf 94}, no. 3, 035008 (2016)
  doi:10.1103/PhysRevD.94.035008
  [arXiv:1603.00360 [hep-ph]].
  
  \bibitem{Zhang:2007fn} 
  Y.~Zhang, H.~An, X.~Ji and R.~N.~Mohapatra,
  Phys.\ Rev.\ D {\bf 76}, 091301 (2007)
  doi:10.1103/PhysRevD.76.091301
  [arXiv:0704.1662 [hep-ph]].
  
  \bibitem{Zhang:2007da} 
  Y.~Zhang, H.~An, X.~Ji and R.~N.~Mohapatra,
  Nucl.\ Phys.\ B {\bf 802}, 247 (2008)
  doi:10.1016/j.nuclphysb.2008.05.019
  [arXiv:0712.4218 [hep-ph]].
  
  \bibitem{Khachatryan:2015dcf} 
  V.~Khachatryan {\it et al.} [CMS Collaboration],
  Phys.\ Rev.\ Lett.\  {\bf 116}, no. 7, 071801 (2016)
  doi:10.1103/PhysRevLett.116.071801
  [arXiv:1512.01224 [hep-ex]].
  
  \bibitem{ATLAS:2015nsi} 
  G.~Aad {\it et al.} [ATLAS Collaboration],
  Phys.\ Lett.\ B {\bf 754}, 302 (2016)
  doi:10.1016/j.physletb.2016.01.032
  [arXiv:1512.01530 [hep-ex]].
  
  \bibitem{ATLAS:2012ak} 
  G.~Aad {\it et al.} [ATLAS Collaboration],
  Eur.\ Phys.\ J.\ C {\bf 72}, 2056 (2012)
  doi:10.1140/epjc/s10052-012-2056-4
  [arXiv:1203.5420 [hep-ex]].
  
  
    \bibitem{Dicus:1991fk} 
  D.~A.~Dicus, D.~D.~Karatas and P.~Roy,
  Phys.\ Rev.\ D {\bf 44}, 2033 (1991).
  doi:10.1103/PhysRevD.44.2033
  
  
  \bibitem{Pilaftsis:1991ug} 
  A.~Pilaftsis,
  Z.\ Phys.\ C {\bf 55}, 275 (1992)
  doi:10.1007/BF01482590
  [hep-ph/9901206].
  
  \bibitem{Datta:1993nm} 
  A.~Datta, M.~Guchait and A.~Pilaftsis,
  Phys.\ Rev.\ D {\bf 50}, 3195 (1994)
  doi:10.1103/PhysRevD.50.3195
  [hep-ph/9311257].
  
  \bibitem{Han:2006ip} 
  T.~Han and B.~Zhang,
  Phys.\ Rev.\ Lett.\  {\bf 97}, 171804 (2006)
  doi:10.1103/PhysRevLett.97.171804
  [hep-ph/0604064].

  \bibitem{Atre:2009rg} 
  A.~Atre, T.~Han, S.~Pascoli and B.~Zhang,
  JHEP {\bf 0905}, 030 (2009)
  doi:10.1088/1126-6708/2009/05/030
  [arXiv:0901.3589 [hep-ph]].
  
  \bibitem{Han:2012vk} 
  T.~Han, I.~Lewis, R.~Ruiz and Z.~g.~Si,
  Phys.\ Rev.\ D {\bf 87}, no. 3, 035011 (2013)
  doi:10.1103/PhysRevD.87.035011
  [arXiv:1211.6447 [hep-ph]].
  
  \bibitem{Chen:2013fna} 
  C.~Y.~Chen, P.~S.~B.~Dev and R.~N.~Mohapatra,
  Phys.\ Rev.\ D {\bf 88}, 033014 (2013)
  doi:10.1103/PhysRevD.88.033014
  [arXiv:1306.2342 [hep-ph]]. 
  
  \bibitem{Dev:2015kca} 
  P.~S.~B.~Dev, D.~Kim and R.~N.~Mohapatra,
  JHEP {\bf 1601}, 118 (2016)
  doi:10.1007/JHEP01(2016)118
  [arXiv:1510.04328 [hep-ph]].
  
  \bibitem{Nemevsek:2012iq} 
  M.~Nemevsek, G.~Senjanovic and V.~Tello,
  Phys.\ Rev.\ Lett.\  {\bf 110}, no. 15, 151802 (2013)
  doi:10.1103/PhysRevLett.110.151802
  [arXiv:1211.2837 [hep-ph]].
  
  \bibitem{Senjanovic:2016vxw} 
  G.~Senjanović and V.~Tello,
  arXiv:1612.05503 [hep-ph].
  
  \bibitem{Khachatryan:2015gha} 
  V.~Khachatryan {\it et al.} [CMS Collaboration],
  Phys.\ Lett.\ B {\bf 748}, 144 (2015)
  doi:10.1016/j.physletb.2015.06.070
  [arXiv:1501.05566 [hep-ex]].
  
  \bibitem{Khachatryan:2016olu} 
  V.~Khachatryan {\it et al.} [CMS Collaboration],
  JHEP {\bf 1604}, 169 (2016)
  doi:10.1007/JHEP04(2016)169
  [arXiv:1603.02248 [hep-ex]].
  
  \bibitem{Alva:2014gxa} 
  D.~Alva, T.~Han and R.~Ruiz,
  JHEP {\bf 1502}, 072 (2015)
  doi:10.1007/JHEP02(2015)072
  [arXiv:1411.7305 [hep-ph]].
  
    \bibitem{delAguila:2008ir} 
  F.~del Aguila, S.~Bar-Shalom, A.~Soni and J.~Wudka,
  Phys.\ Lett.\ B {\bf 670}, 399 (2009)
  doi:10.1016/j.physletb.2008.11.031
  [arXiv:0806.0876 [hep-ph]].
  
  \bibitem{Maiezza:2010ic} 
  A.~Maiezza, M.~Nemevsek, F.~Nesti and G.~Senjanovic,
  Phys.\ Rev.\ D {\bf 82}, 055022 (2010)
  doi:10.1103/PhysRevD.82.055022
  [arXiv:1005.5160 [hep-ph]].
  
  \bibitem{Senjanovic:2014pva} 
  G.~Senjanović and V.~Tello,
  Phys.\ Rev.\ Lett.\  {\bf 114}, no. 7, 071801 (2015)
  doi:10.1103/PhysRevLett.114.071801
  [arXiv:1408.3835 [hep-ph]].
  
  \bibitem{Senjanovic:2015yea} 
  G.~Senjanovic and V.~Tello,
  Phys.\ Rev.\ D {\bf 94}, no. 9, 095023 (2016)
  doi:10.1103/PhysRevD.94.095023
  [arXiv:1502.05704 [hep-ph]].
  

    \bibitem{Aparici:2009fh} 
  A.~Aparici, K.~Kim, A.~Santamaria and J.~Wudka,
  Phys.\ Rev.\ D {\bf 80}, 013010 (2009)
  doi:10.1103/PhysRevD.80.013010
  [arXiv:0904.3244 [hep-ph]].
      
  \bibitem{Bhattacharya:2015vja} 
  S.~Bhattacharya and J.~Wudka,
  Phys.\ Rev.\ D {\bf 94}, no. 5, 055022 (2016)
  Erratum: [Phys.\ Rev.\ D {\bf 95}, no. 3, 039904 (2017)]
  doi:10.1103/PhysRevD.94.055022, 10.1103/PhysRevD.95.039904
  [arXiv:1505.05264 [hep-ph]].
  
  \bibitem{Buchmuller:1985jz} 
  W.~Buchmuller and D.~Wyler,
  Nucl.\ Phys.\ B {\bf 268}, 621 (1986).
  doi:10.1016/0550-3213(86)90262-2
  
  \bibitem{Grzadkowski:2010es} 
  B.~Grzadkowski, M.~Iskrzynski, M.~Misiak and J.~Rosiek,
  JHEP {\bf 1010}, 085 (2010)
  doi:10.1007/JHEP10(2010)085
  [arXiv:1008.4884 [hep-ph]].
  
  \bibitem{Ma:1998dn} 
  E.~Ma,
  Phys.\ Rev.\ Lett.\  {\bf 81}, 1171 (1998)
  doi:10.1103/PhysRevLett.81.1171
  [hep-ph/9805219]. 
   
  \bibitem{Kersten:2007vk} 
  J.~Kersten and A.~Y.~Smirnov,
  Phys.\ Rev.\ D {\bf 76}, 073005 (2007)
  doi:10.1103/PhysRevD.76.073005
  [arXiv:0705.3221 [hep-ph]].  
  
  \bibitem{Minkowski:1977sc} 
  P.~Minkowski,
  Phys.\ Lett.\ B {\bf 67}, 421 (1977).

  \bibitem{Yanagida:1979as} 
  T.~Yanagida,
  Conf.\ Proc.\ C {\bf 7902131}, 95 (1979).

  \bibitem{GellMann:1980vs} 
  M.~Gell-Mann, P.~Ramond and R.~Slansky,
  Conf.\ Proc.\ C {\bf 790927}, 315 (1979)
  
  \bibitem{Mohapatra:1979ia} 
  R.~N.~Mohapatra and G.~Senjanovic,
  Phys.\ Rev.\ Lett.\  {\bf 44}, 912 (1980).
  
  \bibitem{Henning:2015alf} 
  B.~Henning, X.~Lu, T.~Melia and H.~Murayama,
  arXiv:1512.03433 [hep-ph].
  
  \bibitem{Kobach:2016ami}
  A.~Kobach,
  Phys.\ Lett.\ B {\bf 758} (2016) 455
  doi:10.1016/j.physletb.2016.05.050
  [arXiv:1604.05726 [hep-ph]].
  
  \bibitem{Gopalakrishna:2010xm} 
  S.~Gopalakrishna, T.~Han, I.~Lewis, Z.~g.~Si and Y.~F.~Zhou,
  Phys.\ Rev.\ D {\bf 82}, 115020 (2010)
  doi:10.1103/PhysRevD.82.115020
  [arXiv:1008.3508 [hep-ph]].
  
  \bibitem{Harris:2001sx} 
  B.~W.~Harris and J.~F.~Owens,
  Phys.\ Rev.\ D {\bf 65}, 094032 (2002)
  doi:10.1103/PhysRevD.65.094032
  [hep-ph/0102128].
  
  \bibitem{Ruiz:2015zca} 
  R.~Ruiz,
  JHEP {\bf 1512}, 165 (2015)
  doi:10.1007/JHEP12(2015)165
  [arXiv:1509.05416 [hep-ph]].
  
  
  
  \bibitem{Christensen:2008py} 
  N.~D.~Christensen and C.~Duhr,
  \textit{FeynRules - Feynman rules made easy,}
  Comput.\ Phys.\ Commun.\  {\bf 180}, 1614 (2009)
  doi:10.1016/j.cpc.2009.02.018
  [arXiv:0806.4194 [hep-ph]].
  
  \bibitem{Alloul:2013bka} 
  A.~Alloul, N.~D.~Christensen, C.~Degrande, C.~Duhr and B.~Fuks,
  \textit{FeynRules  2.0 - A complete toolbox for tree-level phenomenology,}
  Comput.\ Phys.\ Commun.\  {\bf 185}, 2250 (2014)
  doi:10.1016/j.cpc.2014.04.012
  [arXiv:1310.1921 [hep-ph]].
   
  \bibitem{Degrande:2014vpa} 
  C.~Degrande,
  Comput.\ Phys.\ Commun.\  {\bf 197}, 239 (2015)
  doi:10.1016/j.cpc.2015.08.015
  [arXiv:1406.3030 [hep-ph]].
  
  \bibitem{Alwall:2014hca} 
  J.~Alwall {\it et al.},
  ``The automated computation of tree-level and next-to-leading order differential cross sections, and their matching to parton shower simulations,''
  JHEP {\bf 1407}, 079 (2014)
  doi:10.1007/JHEP07(2014)079
  [arXiv:1405.0301 [hep-ph]].
  
  \bibitem{Read:2002hq} 
  A.~L.~Read,
  J.\ Phys.\ G {\bf 28}, 2693 (2002).
  doi:10.1088/0954-3899/28/10/313
  
  \bibitem{Junk:1999kv} 
  T.~Junk,
  Nucl.\ Instrum.\ Meth.\ A {\bf 434}, 435 (1999)
  doi:10.1016/S0168-9002(99)00498-2
  [hep-ex/9902006].
  
  \bibitem{ATLAS:2011tau} 
  [ATLAS Collaboration],
  ATL-PHYS-PUB-2011-011, ATL-COM-PHYS-2011-818, CMS-NOTE-2011-005.
  
  \bibitem{Hinchliffe:2015qma} 
  I.~Hinchliffe, A.~Kotwal, M.~L.~Mangano, C.~Quigg and L.~T.~Wang,
  Int.\ J.\ Mod.\ Phys.\ A {\bf 30}, no. 23, 1544002 (2015)
  doi:10.1142/S0217751X15440029
  [arXiv:1504.06108 [hep-ph]].
 
\end{thebibliography}
\end{document}